\documentclass[aps,groupedaddress,pra,twocolumn]{revtex4}
\usepackage{graphicx}
\usepackage{enumerate}
\usepackage{amsmath,amssymb,color}
\usepackage{braket}
\usepackage[dvipdfmx,colorlinks,citecolor=red]{hyperref}

\newcommand{\abs}[1]{| #1 |}
\newcommand{\Abs}[1]{\left| #1 \right|}

\newcommand{\Norm}[1]{\left\| #1 \right\|}

\newtheorem{prob}{Problem}
\newtheorem{theo}{Theorem}

\newtheorem{lemm}{Lemma}

\newtheorem{coro}{Corollary}

\begin{document}

\title{A quantum algorithm for 
additive approximation of Ising partition functions}
\author{Akira Matsuo,$^1$ Keisuke Fujii,$^{2,3}$ and Nobuyuki Imoto$^1$}
\date{\today}

\address{$^1$Graduate School of Engineering Science, Osaka University, Toyonaka, Osaka, 560-8531, Japan}
\address{$^2$The Hakubi Center for Advanced Research, Kyoto University, Yoshida-Ushinomiya-cho, Sakyo-ku, Kyoto 606-8302, Japan}
\address{$^3$Graduate School of Informatics, Kyoto University, Yoshida Honmachi, Sakyo-ku, Kyoto 606-8501, Japan}

\begin{abstract}
We investigate quantum computational complexity of 
calculating partition functions of Ising models.
We construct a quantum algorithm for an additive approximation of 
Ising partition functions on square lattices.
To this end, we utilize
the overlap mapping developed by Van den Nest, D{\"u}r, and Briegel [Phys. Rev. Lett. {\bf 98}, 117207 (2007)]
and its interpretation through measurement-based quantum computation (MBQC).
We specify an algorithmic domain,
on which the proposed algorithm works, and an approximation scale,
which determines the accuracy of the approximation.
We show that the proposed algorithm does a nontrivial task,
which would be intractable on any classical computer,
by showing the problem solvable by the proposed quantum algorithm 
are BQP-complete.
In the construction of the BQP-complete problem
coupling strengths and magnetic fields
take complex values.
However, the Ising models that are of central interest in 
statistical physics and computer science consist of real coupling strengths and magnetic fields.
Thus we extend the algorithmic domain of the proposed algorithm 
to such a real physical parameter region
and calculate the approximation scale explicitly.
We found that the overlap mapping and its MBQC interpretation
improves the approximation scale exponentially
compared to a straightforward constant depth quantum algorithm.
On the other hand, 
the proposed quantum algorithm also provides us a partial evidence
that there exist no efficient classical algorithm 
for a multiplicative approximation of the Ising partition functions
even on the square lattice.
This result 
supports that the proposed quantum algorithm
does a nontrivial task also in the physical parameter region.
\end{abstract}

\maketitle
\section{Introduction} \label{sec:I}
Classical spin models
have been widely studied in statistical physics for a long time 
as simplified pictures of magnetic materials.
The Ising model is the simplest model
consisting of two discrete spin variables, up and down,
but exhibits a rich structure
enough to be applied not only for magnetic materials,
but also lattice gases~\cite{LeeYang}, binary alloys, neural systems~\cite{Hopfield1982} and economic models~\cite{Jain2005}. 
One of the main goal is to calculate a partition function, 
which tells us statistical properties of 
a system in thermodynamic equilibrium, such as 
free energy, magnetization, specific heat, and so on. 
Many techniques have been developed 
to calculate the Ising partition functions in both 
exact and approximated manners so far.
Only restricted type of Ising models,
such as Ising models on two-dimensional planer lattices without magnetic 
fields, are exactly solvable~\cite{Kasteleyn1961,Fisher1966}.  
In general, exact calculation of Ising partition functions 
belongs to \#P-hard problems,
which are highly intractable in classical computer~\cite{Barahona1982}.
Furthermore, even an efficient (multiplicative) approximation of 
antiferromagnetic Ising partition functions on general lattices
does not exist unless RP = NP~\cite{Zucherman,RP=NP},
which is highly implausible to occur~\cite{Jerrum1993,Goldberg2008}.
It is a natural question
how quantum computer is useful in this context~\cite{Lidar1997,Master2003}.

Recently quantum information theory sheds new light 
on computational complexity of Ising partition functions.
Bravyi and Raussendorf argued 
classical simulatability of measurement-based quantum computation (MBQC)~\cite{Raussendorf01}
on the planer surface codes by mapping 
it into an Ising partition function on a planer lattice~\cite{Bravyi2007}.
Van den Nest, D{\"u}r, and Briegel
established a correspondence between 
the quantum-stabilizer formalism and classical spin models~\cite{Nest2007}.
In this mapping,
a partition function of a classical spin model
is expressed as an overlap between 
a stabilizer state and a product state.
This overlap mapping allows us to apply powerful results obtained in the context of 
quantum information theory to statistical physics~\cite{Nest2007,Fujii2013rev}.
Although a transfer matrix approach and a state overlap
have already appeared in an earlier work~\cite{Lidar1997},
the overlap mapping makes the problem much more tractable,
allowing us to interpret the overlap as MBQC
and associated quantum circuits.
For example, Van den Nest, D{\"u}r, and Briegel showed that the Ising model on a square lattice 
is complete in the sense that 
a partition function of any classical spin model on an arbitrary graph 
can be expressed as a certain instance of it~\cite{Nest2008,Karimipour}.
Furthermore, classical simulatability of MBQC on certain stabilizer states
provides us an efficient classical algorithm to calculate the corresponding partition functions~\cite{Nest2007,VandenNest2013}.
Based on this mapping,
De las Cuevas {\it et al.}
proposed a quantum algorithm to approximate
partition functions of classical spin models,
such as Ising, Potts, vertex, and gauge models, in a complex parameter regime~\cite{Cuevas2011}.
Furthermore, they showed that
additive approximations of certain classical spin models
are BQP-complete.
(BQP stands for bounded-error quantum polynomial time computation
and is a class of decision problems that can be efficiently solvable
by a quantum computer.)
This means that all problems that are solvable by quantum computer
can be mapped into these problems.
The consequences of this result are twofold.
Firstly, at least for these types of classical spin models,
we can utilize a quantum computer to estimate their partition functions efficiently.
Secondary, BQP-completeness implies that
there is less possibility to do this task on a classical computer.
(If it is possible, we can simulate a quantum computer by a classical computer,
which is highly implausible.)

Besides, quantum computational complexity of not only Ising partition functions
but also link invariants such as Jones and Tutte polynomials 
has been argued also in the circuit models~\cite{Aharonov2006,Aharonov2007,Arad2008,Aharonov2011,Iblisdir2014},
and their additive approximations
have been shown to be BQP-complete.
Recently a sampling problem 
related to Ising partition functions 
has been shown to be intractable on any classical computer,
while it can be done by using commutable quantum circuits~\cite{Fujii2013},
so-called instantaneous quantum polynomial time computation (IQP),
which seems to be much weaker than universal quantum computation~\cite{Bremner2010}.
These results provide a clew to understand 
not only complexity of calculating classical spin models
but also problems that are solvable by a quantum computer and 
the origin of quantum a speedup.

In this paper, we further investigate complexity of 
calculating Ising partition functions 
based on the overlap mapping~\cite{Nest2007} and its interpretation through MBQC.
We specifically consider a quantum algorithm that
approximates Ising partition functions on square lattices,
where each instance of the problem is encoded into the coupling strengths and magnetic fields.
In this sense, 
the present work is complimentary to 
those works done in Ref.~\cite{Nest2009,Cuevas2011}, in which
instances of the problem is encoded into topology of the graphs
taking the coupling strengths and magnetic fields homogeneously.
Furthermore,
we specify a domain, on which the proposed quantum algorithm works,
and an approximation scale, which determines the accuracy of the additive approximation. 
We also provide a proof that 
the problem solved by the proposed quantum algorithm
is BQP-hard.
This indicates that the proposed quantum algorithm
does a nontrivial task, which would be intractable on any classical computer.

We also establish a way to approximate general Ising partition functions
including real coupling strengths and magnetic fields,
which are especially of interest in statistical physics.
In such a physical parameter region,
the gate operations done by MBQC are not unitary anymore,
and linear operators in general.
Thus we construct a quantum algorithm
that approximates such linear operators
following the methods taken in Refs.~\cite{Aharonov2007,Arad2008}.
This allows us to clarify the approximation scale for general coupling strengths and magnetic fields.

We should note that a related work has been done by Iblisdir {\it et al.}
recently~\cite{Iblisdir2014}.
In their work,
certain types of quantum circuits are mapped into 
the Ising partition functions on square lattices.
Specifically, they utilize a transfer matrix approach
to map quantum circuits into the Ising partition functions.
(A similar approach has been also taken in an earlier work~\cite{Lidar1997}.)
The construction of the quantum algorithm in this work can be 
regarded as a measurement-based version of these works,
which would be simpler for people who are familiar with MBQC.
They have also considered
approximation of the Ising partition functions 
with real coupling strengths and magnetic fields,
while a rather different approach, analytic continuation, was taken (see also a related work~\cite{Master2003}).
Instead of analytic continuation,
we here straightforwardly simulate linear operators
by using unitary circuits.
Since the proposed quantum algorithm
provides the approximation scale explicitly for all parameter region,
the proposed quantum algorithm allows us to compare the performances.
Furthermore, 
the quantum circuits that approximate the physical Ising partition functions 
are also provide explicitly,
which would be helpful to understand performance of the 
proposed quantum algorithm in the physical parameter region.

Unfortunately, it is still unknown 
whether or not the proposed quantum algorithm 
does a nontrivial task in the physical parameter region.
However,
we also provides a partial evidence that
a multiplicative approximation of the Ising partition functions
in the physical parameter region
cannot be attained by any classical computer
unless the polynomial hierarchy collapses at the third level.
This result strongly supports that the proposed quantum algorithm
actually does a nontrivial task even in the physical parameter region.
We believe these quantum algorithms in 
the real parameter regime and their approximation scales
provide an essential clue to understand potential of quantum computation
in solving problems that takes an important role in combinatorial optimization problems.

The rest of the paper is organized as follows.
In Sec.~\ref{sec:II},
we review the correspondence between
the quantum stabilizer formalism and the Ising partition functions
with fixing the notations.
In Sec.~\ref{sec:III},
we propose a quantum algorithm that 
approximates the Ising partition functions on square lattices.
We also show that the proposed quantum algorithm solves
a BQP-complete problem and hence does a nontrivial task,
which would be intractable on any classical computer.
In Sec.~\ref{sec:IV},
we extend the domain of the 
proposed quantum algorithm to general coupling strengths and magnetic fields,
which include real parameters.
We calculate the approximation scale 
of the quantum algorithm in this domain
and argue the performance of the proposed quantum algorithm
in the physical parameter region.
Section~\ref{sec:V} is devoted to conclusions and discussions.

\section{Quantum formulation of Ising model} \label{sec:II}
In this section, we briefly review 
the correspondence between the quantum-stabilizer formalism and 
Ising partition functions with fixing the notations.
We consider a classical Ising model defined on a graph $G=(V,E)$ with 
$V$ and $E$ being sets of vertices and edges, respectively.
The Ising model consists of classical two-state spin variables $\sigma_a=\pm 1$
defined on each vertex $a\in V$ of the graph $G$. 
Two spins $\sigma _a$ and $\sigma _b$ connected by an edge $\{a,b\}\in E$ 
interact with each other by a coupling strength $J_{ab}$.
Furthermore, each spin $\sigma _a$ is
subjected to a local magnetic field $h_a$.
The Hamiltonian of the system is given by 
\begin{equation}
H_G(\mbox{\boldmath $\sigma$})=-\sum_{\{ a,b\}\in E}J_{ab}\sigma_a\sigma_b-\sum_{a\in V}h_a\sigma_a,
\end{equation}
where $\mbox{\boldmath $\sigma$}$ indicates a spin configuration.
The partition function is defined by 
\begin{equation}
Z_G=\sum_{\mbox{\boldmath $\sigma$}}e^{-\beta H_G(\mbox{\boldmath $\sigma$})},
\end{equation} 
where the summation $\sum _{\mbox{\boldmath $\sigma$}}$ is taken over all spin configurations,
and $\beta$ is the inverse temperature, $\beta=1/k_\mathrm{B}T$, with $k_\mathrm{B}$ and $T$ being the Boltzmann constant and the temperature, respectively.

\begin{figure*}
\centering
\includegraphics[width=170mm]{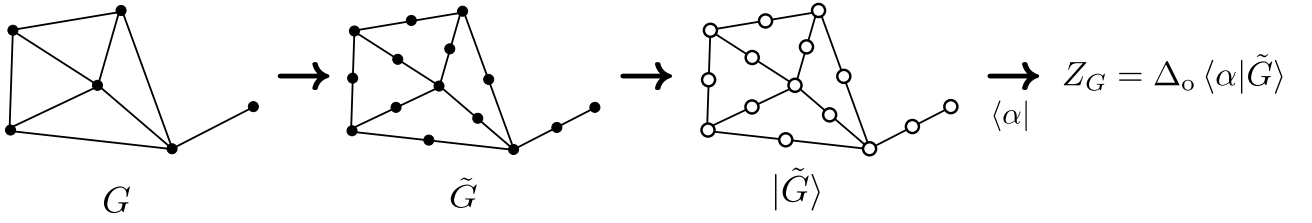}
\caption{
The graph $G$, the corresponding decorated graph $\tilde{G}$,
and the decorated graph state $|\tilde G\rangle$ (from the left to the right).
The partition function of the Ising model on a graph $G$ is 
related with an inner product between the graph state $\ket{\tilde{G}}$ and a product state $\ket{\alpha}$ defined in Eqs. \eqref{eq:6}-\eqref{eq:8} with an approximation scale $\Delta_{\mathrm{o}}$.
}
\label{fig:decorated graph}
\end{figure*}
We express the partition function as an inner product 
between a product state and a stabilizer state~\cite{Nest2008}.
We first define the stabilizer state,
which is described as a graph state associated with another graph $\tilde G$.
The graph $\tilde G=(\tilde V, \tilde E)$, which we call a decorated graph,
is defined by adding a vertex on each edge of the graph $G$
as shown in Fig.~\ref{fig:decorated graph}.
The decorated graph $\tilde G$ has $\abs{\tilde{V}}=|V|+|E|$ vertices and $\abs{\tilde{E}}=2|E|$ edges.
The set of vertices $\tilde{V}$ is defined by $\tilde V =V \cup V_E$, where $V_E=\set{ e_{ab}|\{ a,b\}\in E}$ corresponds to 
the set of vertices added on the edges.
The set of edges $\tilde{E}$ is defined by $\tilde{E}=\set{\{a,e_{ab}\}|a,b\in V, e_{ab}\in V_E}$.
Assigning a qubit on each vertex in $\tilde{V}$, 
we define a $\abs{\tilde{V}}$-qubit stabilizer state:
\begin{equation}
\ket{\varphi_{\tilde{G}}} = 2^{-|V|/2}\sum_{\mbox{\boldmath $s$}} \bigotimes_{e_{ab}\in V_E}\ket{s_a\oplus s_b}\bigotimes_{a\in V}\ket{s_a},
\end{equation}
where $s_a=0, 1$, $\sum_{\mbox{\boldmath $s$}}$ is taken over all configurations of $\mbox{\boldmath $s$}\equiv \{ s_a\}$,
and $s_a\oplus s_b$ denotes the addition modulo 2.
The binary variable $s_a=0,1$ is related with the Ising spin $\sigma _a=\pm 1$ later.
The qubits belonging to $V$ and $V_E$ are referred to as vertex and edge qubits, respectively.
We can easily confirm $\ket{\varphi_{\tilde{G}}}$ is a stabilizer state,
since it can be obtained as
\begin{equation}
\ket{\varphi_{\tilde{G}}} = \left[ \prod _{a \in V}\prod_{e\in \tilde{\mathcal{N}}_a} \Lambda _{a,e}(X) \right]
\bigotimes_{a\in V}\ket{+}\bigotimes_{e\in V_E}\ket{0},
\end{equation}
where $\Lambda _{i,j} (A)$ indicates the controlled-$A$ gate between 
qubit $i$ (control) and $j$ (target),
and $\tilde{\mathcal{N}}_a\subseteq V_E$ denotes the set of vertices adjacent to vertex $a\in V$ on the decorated graph $\tilde G$.
Let $\ket{\tilde{G}}$ be the graph state associated with the graph $\tilde{G}$~\cite{Hein2006}.
Using the equality $\Lambda _{i,j} (X) = H_j \Lambda _{i,j}(Z) H_j$,
$\ket{\varphi_{\tilde{G}}}$ is related with the graph state $\ket{\tilde{G}}$ as follows:
\begin{align}
\ket{\varphi_{\tilde{G}}} 
&= 
\left( \prod_{e_{ab}\in V_E} H_{e_{ab}} \right)
\left[ \prod _{a \in V} \prod _{e \in \tilde{\mathcal{N}}_a} \Lambda _{a,e}(Z) \right]
\ket{+}^{\otimes |V|+|V_E|}
\nonumber\\
&=
\left( \prod_{e_{ab}\in V_E} H_{e_{ab}} \right) \ket{\tilde{G}},
\end{align}
where $H_{e_{ab}}$ is the Hadamard gate acting on edge qubit on $e_{ab}$.

Next we define a product state with which 
the stabilizer state $\ket{\varphi_{\tilde{G}}}$ is taken an inner product:
\begin{eqnarray}
\ket{\alpha}=\bigotimes_{e_{ab}\in V_E}H\ket{\alpha_{e_{ab}}}\bigotimes_{a\in V}\ket{\alpha_a},
\label{eq:6}
\end{eqnarray}
where the single-qubit states are defined as
\begin{align}
\bra{\alpha_{e_{ab}}} &= \frac{e^{\beta J_{ab}}\bra{0}_{e_{ab}}+e^{-\beta J_{ab}}\bra{1}_{e_{ab}}}{\sqrt{\Abs{e^{\beta J_{ab}}}^2+\Abs{e^{-\beta J_{ab}}}^2}},  
\label{eq:7}
\\
\bra{\alpha_{a}} &= \frac{e^{\beta h_a}\bra{0}_{a}+e^{-\beta h_a}\bra{1}_{a}}{\sqrt{\Abs{e^{\beta h_{a}}}^2+\Abs{e^{-\beta h_{a}}}^2}},
\label{eq:8}
\end{align}
for all vertices $e_{ab}\in V_E$ and $a\in V$.

Now we relate
the Ising partition function with the inner product between
the product state $|\alpha\rangle$ and the graph state $\ket{\tilde{G}}$
as follows:
\begin{align}
Z_G &=
\Delta_\mathrm{o}
\left(\bigotimes _{e_{ab} \in V_E} \langle \alpha _{e_{ab}} |
\bigotimes _{a \in V} \langle \alpha _{a} | \right) | \varphi _{\tilde G}\rangle 
\nonumber \\
&=
\Delta_\mathrm{o}\braket{\alpha| \tilde{G}},
\label{eq:partition function overlap}
\end{align}
where the approximation scale $\Delta_\mathrm{o}$ is defined by
\begin{align}\label{eq:Delta overlap}
\Delta_\mathrm{o}&=
2^{|V|/2}\prod_{\{a,b\}\in E}\sqrt{\Abs{e^{\beta J_{ab}}}^2+\Abs{e^{-\beta J_{ab}}}^2}  \nonumber \\
&\qquad \times\prod_{a\in V}\sqrt{\Abs{e^{\beta h_{a}}}^2+\Abs{e^{-\beta h_{a}}}^2}.
\end{align}
Eq.~\eqref{eq:partition function overlap}
can be understood as follows.
The stabilizer state
$\ket{\varphi_{\tilde{G}}}$ (or the graph state $|\tilde G \rangle$)
has information of the geometry of the Ising interactions.
More precisely, each vertex qubit has a superposition of spin up and down states,
and each edge qubit encodes the information 
whether the two spins interacting with are parallel or antiparallel.
Depending on the state of the vertex and edge qubits,
weights $e^{ \pm \beta h_{a}}$ and $e^{ \pm \beta J_{ab}}$ are assigned 
by the product state through the inner product.
Then,
the superposition of all spin states is reduced to
the summation over all spin configuration,
which recovers the partition functions $Z_{G}$
in the l.h.s. of Eq. \eqref{eq:partition function overlap}.

\begin{figure*}
\centering
\includegraphics[width=140mm]{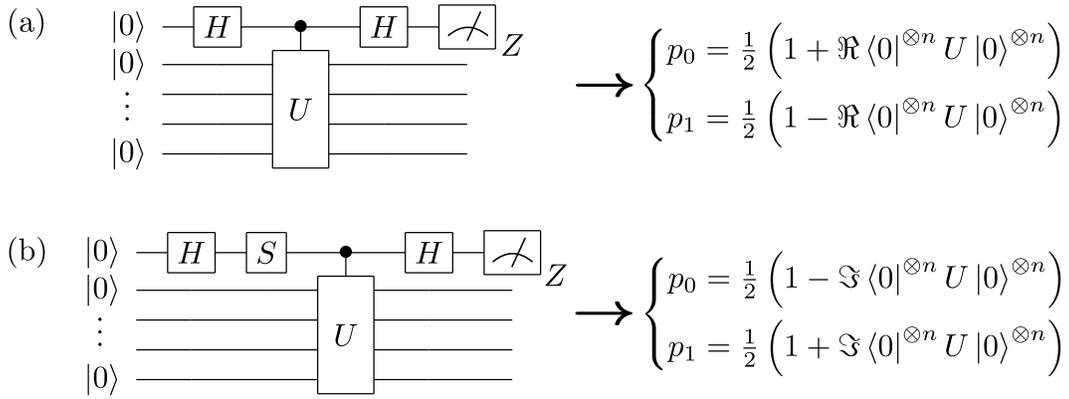}
\caption{The Hadamard test to estimate a matrix element 
$\bra{0}^{\otimes n}U\ket{0}^{\otimes n}$ with an additive error $1/\mathrm{poly}(n)$.
The real and imaginary parts of the matrix element,
$\Re\left(\bra{0}^{\otimes n}U\ket{0}^{\otimes n}\right)$ and $\Im\left(\bra{0}^{\otimes n}U\ket{0}^{\otimes n}\right)$ respectively,
are estimated from the probability distributions 
of the $Z$-basis measurements in the circuits (a) and (b), respectively.
Here $\Re(\cdot)$ and $\Im(\cdot)$ indicate the real and imaginary parts, respectively.
}
\label{fig:Hadamard test}
\end{figure*}

Next we translate the overlap Eq.~\eqref{eq:partition function overlap} 
into quantum computation, which is one of the most important
task to establish a bridge between 
quantum computation and Ising partition functions.
The states $\bra{\alpha_{e_{ab}}}$ and $\bra{\alpha_a}$ can be expressed by using unitary gates $A_{e_{ab}}$ and $A_a$ acting on the computational basis state respectively:
\begin{equation}
\bra{\alpha_{e_{ab}}}=\bra{0}A_{e_{ab}}, \;\;\; \bra{\alpha_a}=\bra{0}A_{a},
\end{equation}
where we defined unitary gates
\begin{align}
A_{e_{ab}}&=\frac{1}{\sqrt{\Abs{e^{\beta J_{ab}}}^2+\Abs{e^{-\beta J_{ab}}}^2}}
\begin{pmatrix}
e^{\beta J_{ab}} & e^{-\beta J_{ab}}  \\
(e^{-\beta J_{ab}})^\ast & -(e^{\beta J_{ab}})^\ast
\end{pmatrix}, \\
A_{a}&=\frac{1}{\sqrt{\Abs{e^{\beta h_{a}}}^2+\Abs{e^{-\beta h_{a}}}^2}}
\begin{pmatrix}
e^{\beta h_a} & e^{-\beta h_a}  \\
(e^{-\beta h_a})^\ast & -(e^{\beta h_a})^\ast
\end{pmatrix}.
\end{align}
Then, the product state $\bra{\alpha}$ can be rewritten as
\begin{align}
\bra{\alpha}&=\bra{0}^{\otimes |V|+|E|}
\left(\bigotimes_{e_{ab}\in V_E}A_{e_{ab}}H\right)
\left(\bigotimes_{a\in V}A_a\right) 
\nonumber \\
&\equiv \bra{0}^{\otimes|\tilde{V}|}A,
\end{align}
where $A$ is defined as a tensor product of single-qubit gates.
On the other hand, by virtue of the properties of the graph state~\cite{Hein2004},
there exists a $|\tilde{V}|$-qubit unitary gate $F$ such that
\begin{equation}
\ket{\tilde{G}}=F\ket{0}^{\otimes |\tilde{V}|}.
\end{equation}
Specifically, if the degree of the graph $\tilde G$ is finite as considered here, 
$F$ is a constant depth Clifford circuit
consisting of Hadamard gates and controlled-$Z$ gates.
Then Eq.~\eqref{eq:partition function overlap} 
is rewritten as
\begin{equation}\label{eq:pf overlap Hadamard test}
Z_G=\Delta_\mathrm{o}\bra{0}^{\otimes|\tilde{V}|}AF\ket{0}^{\otimes|\tilde{V}|}.
\end{equation}
The quantum circuit $AF$ consisting of only $\mathrm{poly}(|\tilde{V}|)$ quantum gates
can be efficiently implemented on a quantum computer.
The matrix element of $AF$ can be estimated
by using the Hadamard test (see e.g.~\cite{Arad2008, Aharonov2006, Aharonov2007}) 
as shown in Fig.~\ref{fig:Hadamard test}.
More precisely, 
we can obtain an approximation $c$ 
of $\bra{0}^{\otimes|\tilde{V}|}AF\ket{0}^{\otimes|\tilde{V}|}$ 
within the following additive error:
\begin{equation}
\Abs{c-\bra{0}^{\otimes|\tilde{V}|}AF\ket{0}^{\otimes|\tilde{V}|}}\le\frac{1}{\mathrm{poly}(|\tilde{V}|)}.
\end{equation}
Accordingly 
the partition function of Eq.~\eqref{eq:partition function overlap} 
can be efficiently approximated with an additive error $\Delta_{\mathrm{o}}/\mathrm{poly}(|\tilde{V}|)$.

The approximation scale $\Delta_{\mathrm{o}}$ of the above quantum algorithm is far from optimal,
since we utilized only constant depth quantum circuits.
By using the idea of MBQC,
we can compress the number of qubits employed 
utilizing non-constant depth quantum circuits.
This allows us to improve the approximation scale as follows.
The overlap $\langle \alpha | \tilde G \rangle$ is 
regarded as an MBQC 
implemented by the sequence of projections $\langle \alpha |$ 
on the resource state $| \tilde G \rangle$.
If the projection $\langle \alpha|$ satisfies 
a certain condition such that the MBQC interpretation works appropriately, 
we can rewrite the overlap as teleportation-based $n$-qubit quantum computation
\begin{eqnarray}
\langle \alpha | {\tilde G} \rangle = 2^{ -(|\tilde V| -n)/2} 
\langle 0|^{\otimes n} U| 0 \rangle^{\otimes n} ,
\end{eqnarray}
where $U$ is a non-constant depth quantum circuit.
Note that the number $n$ of qubits in the r.h.s.\ are reduced compared to 
that $|\tilde V|$ in the l.h.s.
By using this identity,
Eq.~\eqref{eq:partition function overlap} can be rewritten as
\begin{eqnarray}
Z_{G} = \Delta \langle 0|^{\otimes n} U| 0 \rangle^{\otimes n}, 
\end{eqnarray}
where an approximation scale is defined as 
$\Delta \equiv \Delta _{\mathrm{o}}2^{ -(|\tilde V| -n)/2} $.
By using the hadamard test to evaluate the r.h.s.,
the partition function $Z_G$ is approximated with an additive error $\Delta /{\rm poly}(n)$.
Note that the approximation scale $\Delta$ is exponentially improved from $\Delta _{\mathrm{o}}$.
On the other hand, computation time is increased only polynomially.
Thus we can still obtain an exponential improvement of the accuracy of the approximation
taking into account the computation time.
A detail of embedding the Ising models in MBQC tells us 
the improved approximation scale and the algorithmic domain
as seen below.

\section{A quantum algorithm for Ising partition functions} \label{sec:III}
In this section, we propose a quantum algorithm to approximate 
the partition function of the Ising model by establishing a mapping
between a class of Ising models
and MBQC.

\begin{figure}
\centering
\includegraphics[width=85mm]{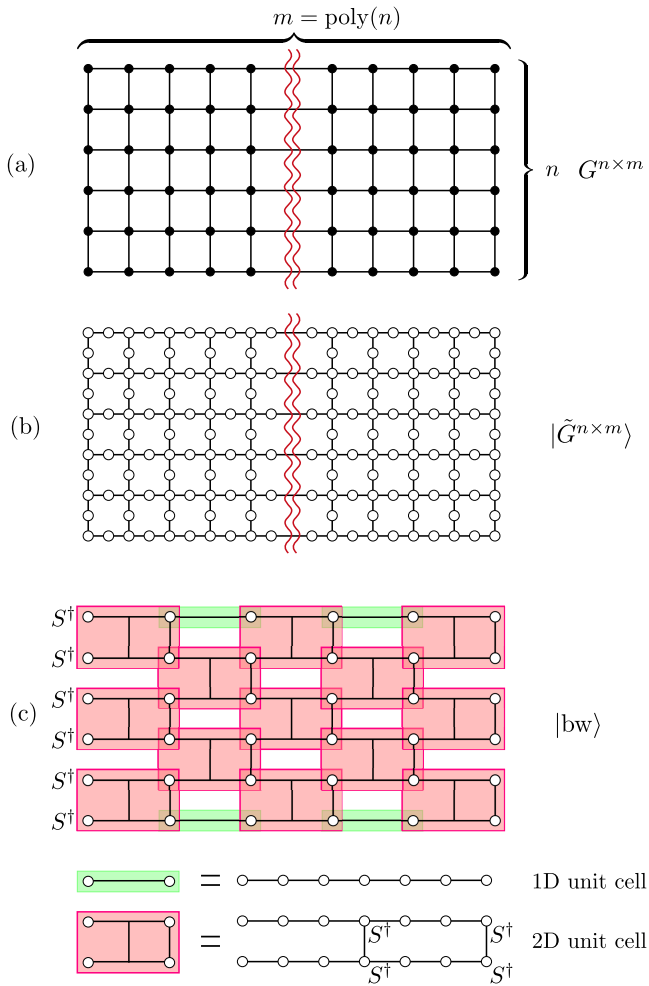}
\caption{
(color online)
(a) The graph $G^{n\times m}$ on which the Ising model is defined.
(b) The corresponding decorated graph state $\ket{{\tilde{G}^{n\times m}}}$.
(c) The brickwork state $\ket{\mathrm{bw}}$
consisting of two-dimensional (2D) unit cells and one-dimensional (1D) unit cells at the top and bottom boundaries.
}
\label{fig:brickwork state}
\end{figure}

We consider an Ising model on an $n\times m$ square lattice $G^{n\times m}$ with $m=\mathrm{poly}(n)$ [see Fig.~\ref{fig:brickwork state} (a)].
We define vertical and horizontal coupling strengths,
$J^{\mathrm{v}}_{ab}$ and $J^{\mathrm{h}}_{ab}$,
for the vertical and horizontal edges $\{a,b\}$, respectively.
The Hamiltonian is given by
\begin{align}
&H_{G^{n\times m}}(\mbox{\boldmath $\sigma$})
\nonumber 
\\
&= -\sum_{\{a,b\}\in E^\mathrm{v}}J_{ab}^\mathrm{v}\sigma_a\sigma_b
-\sum_{\{a,b\}\in E^\mathrm{h}}J_{ab}^\mathrm{h}\sigma_a\sigma_b
-\sum_{a\in V}h_a\sigma_a,
\end{align}
where $E^{\mathrm{v}}$ and $E^{\mathrm{h}}$ are the sets of vertical and horizontal edges, respectively.
Specifically, we consider the following problem:
\begin{prob}
\label{prob1}
({\bf Approximation of Ising partition functions})
Consider an Ising model on an $n\times m$ square lattice, where $m=\mathrm{poly}(n)$.
The magnetic fields $\{\beta h_a \}$ and the vertical coupling strengths $\{\beta J^{\mathrm{v}}_{ab}\}$
are arbitrary imaginary numbers and the horizontal coupling strengths $\{\beta J^{\mathrm{h}}_{ab}\}$
are given by $\{ r_{ab} + i(2k _{ab}+1) \pi/4 \}$,
where $r_{ab}$ is a real number and $k_{ab}$ is an integer.
The problem is defined as an approximation of 
the partition function $Z_{G^{n \times m}}$ of the given Ising Hamiltonian
$H_{G^{n\times m}}$ with an additive error 
$\Delta /{\rm poly}(n)$,
where the approximation scale $\Delta$ is given by $\Delta =2^{n(m+1)/2} \prod _{\{a,b\} \in E^{\mathrm{h}}} \sqrt{\cosh(2r_{ab})}$.
\end{prob}

In the following subsections, we will show two theorems:
\begin{theo}[Quantum algorithm]
\label{the:algo}
There exists an efficient quantum algorithm that solves Problem~\ref{prob1}.
\end{theo}
\begin{theo}[BQP-hardness]
Problem~\ref{prob1} is BQP-hard.
\label{the:hard}
\end{theo}
By combining these results,
we will conclude the following theorem:
\begin{theo}[BQP-completeness]
\label{the:comp}
Consider Problem~\ref{prob1} and $\Abs{Z_{G^{n \times m}}}$ is promised to be either $\le \Delta/3$ or $\ge 2\Delta/3$.
Then the problem to decide whether $\Abs{Z_{G^{n \times m}}}\le \Delta/3$ or not is BQP-complete.
\end{theo}

\subsection{Construction of quantum algorithm\\ (proof of Theorem~\ref{the:algo})}
\begin{figure*}
\centering
\includegraphics[width=160mm]{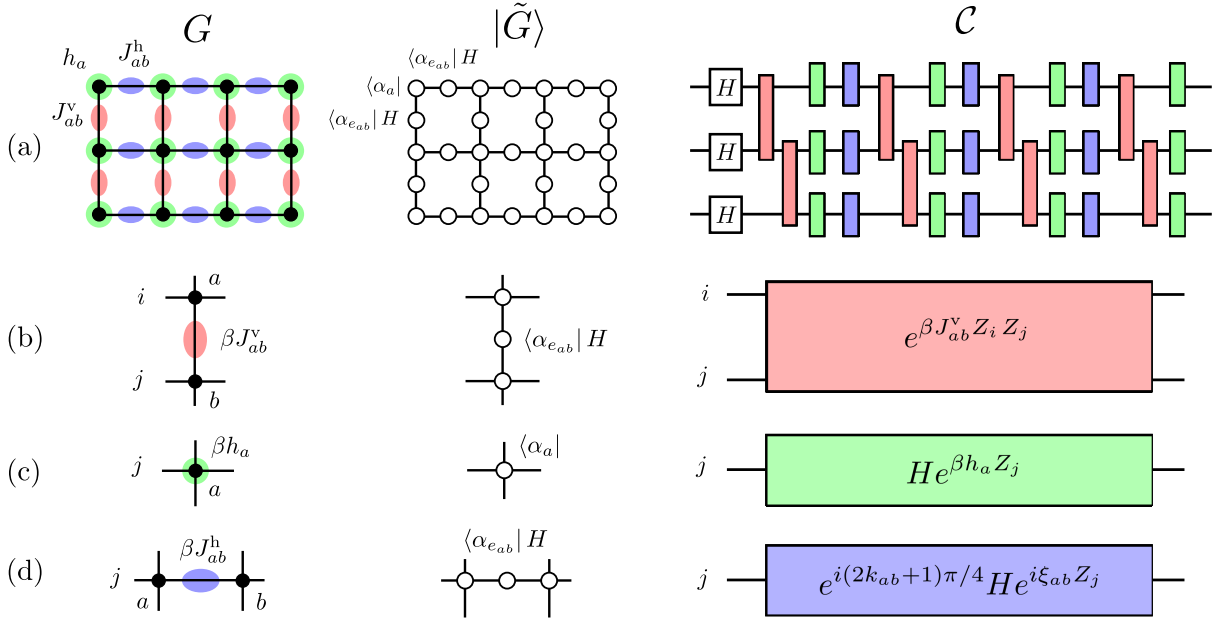}
\caption{
(color online)
The left figures show the parameters of the Ising model. 
The middle figures show the projections made on the 
graph state.
The right figures show the resulting quantum gates.
(a) The whole pictures of constructing the quantum circuits.
(b) The vertical coupling strength and the resulting 
two-qubit gate.
(c) The magnetic field and the resulting single-qubit gate.
(d) The horizontal coupling strength and the resulting single-qubit gate.
}
\label{fig:algorithm for circuit}
\end{figure*}
In this subsection, we prove Theorem~\ref{the:algo}.
We first construct a quantum algorithm which 
solves Problem~\ref{prob1}.
To this end,
we interpret horizontal edges of the graph $G^{n\times m}$ as wires of an $n$-qubit quantum circuit.
According to the coupling strengths and magnetic fields,
quantum gates are assigned on the wires from the left to the right 
as follow [see Fig.~\ref{fig:algorithm for circuit} (a)]:
\begin{enumerate}[(i)]
\item $H^{\otimes n}$ are assigned as initial gates.

\item\label{step1} According to the vertical coupling strength $\beta J_{ab}^{\mathrm{v}}$,
A two-qubit gate
$U_{e_{ab}}\equiv e^{\beta J_{ab}^{\mathrm{v}} Z_i Z_j}$ is assigned on the corresponding $i$th and $j$th wires.
Since $\beta J_{ab}^{\mathrm{v}}$ is an imaginary number,
$U_{e_{ab}}$ is a two-qubit unitary gate.
Specifically, if $\beta J_{ab}^{\mathrm{v}}=0$, an identity gate is assigned.

\item According to the magnetic field $\beta h_a$,
a single-qubit gate $U_a \equiv He^{\beta h_a Z_j}$
is assigned on the corresponding $j$th wire. 
Since $\beta h_a$ is an imaginary number, 
$U_a$ is a single qubit unitary gate.

\item\label{step2} According to the horizontal coupling strength $\beta J_{ab}^{\mathrm{h}}$,
we assign a single-qubit gate
$U_{e_{ab}}\equiv e^{i(2k_{ab}+1)\pi/4}He^{i \xi_{ab} Z_j}$ on the corresponding $j$th wire.
Here $\xi_{ab} \in [-\pi/2,\pi/2]$ is an angle that satisfies
$\sin\xi_{ab}=(-1)^{k_{ab}+1}e^{-r_{ab}}/\sqrt{2\cosh(2r_{ab})}$. 
[Recall that $\beta J_{ab}^{\mathrm{h}} = r_{ab} + i (2 k_{ab} +1)\pi/4$.]

\item Repeat the steps (\ref{step1})-(\ref{step2}) 
for each column of the decorated graph $\tilde G^{n \times m}$
from the left to right. 

\end{enumerate}
After the above procedure, we obtain a quantum circuit 
\begin{eqnarray}
\mathcal{C} = \left(\prod _{\eta \in \tilde V^{n \times m}}^{\to} U_{\eta } \right) H^{ \otimes n},
\end{eqnarray}
where the multiplication $\vec{\prod} _{\eta \in \tilde V^{n \times m}}$ is performed over 
all vertices $\eta \in \tilde V^{n \times m}$ of the decorated graph $\tilde G^{n \times m}$ 
from the left to right columns.
In the same column, the multiplications for vertical edge qubits are taken at first,
and then those for vertex qubits are taken secondarily. 

Next we will show
that the quantum circuit $\mathcal{C}$ is related to the partition function as
\begin{eqnarray}
Z_{G^{n\times m}} \propto \langle \alpha | \tilde G^{n \times m} \rangle\propto \bra{0}^{\otimes n}\mathcal{C}\ket{0}^{\otimes n}.
\end{eqnarray}
We interpret the projection by $\langle \alpha|$
as a sequence of measurements in MBQC, 
whose resource state is given by the graph state $\ket{{\tilde{G}^{n\times m}}}$.
The measurements are assumed to be performed 
from the left to the right.
(In the projection $\langle \alpha |$,
the measurement outcomes are always determined, and
there is no feedforward in the present MBQC interpretation.
Thus we chose a convenient measurement order, 
from the left to the right, without loss of generality.)
As shown below, the projection by $\langle \alpha |$ induces
a sequence of unitary gates in a measurement-based way,
which corresponds to the $n$-qubit quantum circuit $\mathcal{C}$ constructed.

The projection 
$\langle \alpha _{e_{ab}} |H = (e^{ \beta J _{ab}^{\mathrm{v}}} \langle 0| + e^{-\beta J _{ab}^{\mathrm{v}}}\langle 1| )H/\sqrt{2}$
on the vertical edge qubit can be written as
\begin{align}
&\langle \alpha _{e_{ab}} | H_{e_{ab}} \Lambda _{a,e_{ab}} (Z) 
\Lambda _{b,e_{ab}} (Z) |+\rangle _{e_{ab}} | {\tilde G^{ n \times m} }\backslash e_{ab} \rangle 
\nonumber \\
&=
\frac{1}{\sqrt{2}} U_{e_{ab}} |  {\tilde G^{ n \times m} }\backslash e_{ab} \rangle.
\end{align}
Here $|{\tilde G^{ n \times m} }\backslash e_{ab} \rangle$ indicates 
a graph state associated with the decorated graph, 
where vertex $e_{ab}$ and adjacent edges are deleted.
This tells us that the projection on the vertical edge qubit on $e_{ab}$
can be replaced by the two-qubit gate $U_{e_{ab}}= e^{\beta J_{ab}^{\mathrm{v}} Z_i  Z_j}$ on the corresponding wires
as shown in Fig.~\ref{fig:algorithm for circuit} (b).

The projection on the vertex qubit
is regarded as quantum teleportation,
which propagates quantum information from the left to the right 
with a single-qubit unitary gate.
The standard argument for MBQC~\cite{Raussendorf01} tells us
the projection by $\langle \alpha _{a}| = ( e^{ \beta h_a} \langle 0|+ e^{-\beta h_a}\langle 1| )/\sqrt{2}$ results in the single-qubit gate $U_{a}=He^{\beta h_a Z_j}$.
The projection on the horizontal edge qubit 
is done with the Hadamard gate:
\begin{align}
&\langle \alpha _{e_{ab}} |H  
\nonumber \\
&= 
\left[ \cosh(\beta J_{ab}^{\mathrm{h}}) \langle 0 |
+ \sinh(\beta J_{ab}^{\mathrm{h}}) \langle 1| \right]
/\sqrt{\cosh (2 r_{ab})}
\nonumber \\
&= e^{i(2k_{ab}+1)\pi/4}
\left( e^{ i \xi_{ab}}\langle 0| + e^{- i \xi_{ab}} \langle 1| \right)/\sqrt{2}.
\end{align}
Recall that $\xi_{ab} \in[-\pi/2,\pi/2]$ and satisfies that 
$\sin\xi_{ab}=(-1)^{k_{ab}+1}e^{-r_{ab}}/\sqrt{2\cosh(2r_{ab})}$.
Similarly to the previous case,
this projection results in a single-qubit gate $U_{e_{ab}}=e^{i(2k_{ab}+1)\pi/4}He^{i \xi_{ab} Z_j}$.

By repeatedly using the above arguments, 
we obtain a unitary gate 
\begin{eqnarray}
\prod _{a\in V_r}  U_{a}^{\dag} \left(\prod _{\eta \in \tilde V^{n \times m}}^{\to} U_{\eta } \right), 
\end{eqnarray}
where $V_r$ is the set of $n$ vertices of the right boundary of $G$.
The initial state of MBQC is $|+\rangle ^{ \otimes n} = H^{\otimes n} |0\rangle ^{ \otimes n}$, 
and hence the Hadamard gates
are implemented on $|0\rangle ^{ \otimes n}$ as initial gates.
The readout of the output qubits are 
done by the projections $\bigotimes_{a\in V_r}\langle \alpha _a| = \bigotimes_{a\in V_r}(\langle 0 |A_a)$ on vertex qubits at the right boundary.
Since $A_a=U_a=He^{\beta h_a Z_j}$,
$U_{a}^{\dag}$ and $A_a$ are canceled out.
This yields the following relation
\begin{equation}
Z_{G^{n\times m}} = \Delta _{\mathrm{o}} \langle \alpha | {\tilde G^{n \times m}} \rangle\propto \bra{0}^{\otimes n}\mathcal{C}\ket{0}^{\otimes n},
\end{equation}
where 
$\Delta _{\mathrm{o}} = 2^{|V|+|V_E|/2} \prod _{\{a,b\} \in E^{\mathrm{h}}} \sqrt{\cosh(2r_{ab})}$.
Since the probability amplitude of the resource state is reduced by $2^{-1/2}$ for each projection in MBQC,
we obtain 
\begin{eqnarray}
\langle \alpha | {\tilde G^{n \times m}} \rangle = 2^{- (|V|+|V_E|-n)/2} \bra{0}^{\otimes n}\mathcal{C}\ket{0}^{\otimes n}.
\end{eqnarray}
Thus we conclude that 
\begin{eqnarray}
Z_{G^{n\times m}} = \Delta \bra{0}^{\otimes n}\mathcal{C}\ket{0}^{\otimes n},
\label{eq:hoge}
\end{eqnarray}
where we defined the approximation scale 
\begin{align}
\Delta
&\equiv \Delta _{\mathrm{o}} 2^{- (|V|+|V_E|-n)/2}  \nonumber \\
&= 2^{|V|/2+ n/2} \prod _{\{a,b\} \in E^{\mathrm{h}}} \sqrt{\cosh(2r_{ab})}.
\end{align}
The matrix element $\bra{0}^{\otimes n}\mathcal{C}\ket{0}^{\otimes n}$
can be estimated with an additive error $1/{\rm poly}(n)$ using the Hadamard test (see Fig.~\ref{fig:Hadamard test}),
which utilizes the controlled-$\mathcal{C}$ gate.
Accordingly the quantum algorithm consisting of 
the $(n+1)$-qubit controlled-$\mathcal{C}$ gate
and the single-qubit measurement for the Hadamard test
approximates the partition function $Z_{G^{n\times m}}$
with an additive error $\Delta /{\rm poly}(n)$.
While the approximation scale is improved exponentially
compared to that $\Delta _{\mathrm{o}}$ for the constant depth algorithm,
it is still unclear whether or not the constructed quantum algorithm 
do a nontrivial task, which would be intractable on any classical computer.
To provide such an evidence,
in the next subsection, we will show 
Problem~\ref{prob1} is BQP-hard (Theorem~\ref{the:hard}).
That is, we can simulate an arbitrary quantum computation
by calculating a partition function with a specific parameter in Problem~\ref{prob1}.

\subsection{BQP-hardness\\ (proof of Theorem~\ref{the:hard})}
In this subsection,
we will prove Theorem~\ref{the:hard},
that is, BQP-hardness of Problem~\ref{prob1}.
We define a subproblem of Problem~\ref{prob1}:
\begin{prob}[BQP-hard subproblem]
\label{prob2}
Consider an Ising model on $G^{n \times m}$ and the corresponding decorated graph state $|\tilde G^{n\times m}\rangle$.
The magnetic fields are taken homogeneously as $\beta h_{a}=i\pi/4$. 
The vertical coupling strengths $\{ \beta J^{\mathrm{v}}_{ab}\}$
are chosen to be $0$ or $i\pi/4$. 
The horizontal coupling strengths $\{ \beta J^{\mathrm{h}}_{ab}\}$
are chosen to be $i\pi/4$ or $ \Omega \equiv \ln(\sqrt{2}+1)/2+i\pi/4$.
Then the problem is defined as an approximation of 
the partition function $Z_{G^{n \times m}}$ of the given Ising Hamiltonian
$H_{G^{n\times m}}$ with an additive error 
$\Delta /{\rm poly}(n)$.
The approximation scale is defined to be $\Delta =2^{n(m+1)/2+\#\Omega/4}$
with $\#\Omega$ being the number of the horizontal couplings of
$\Omega$.

\end{prob}

\begin{figure}
\centering
\includegraphics[width=85mm]{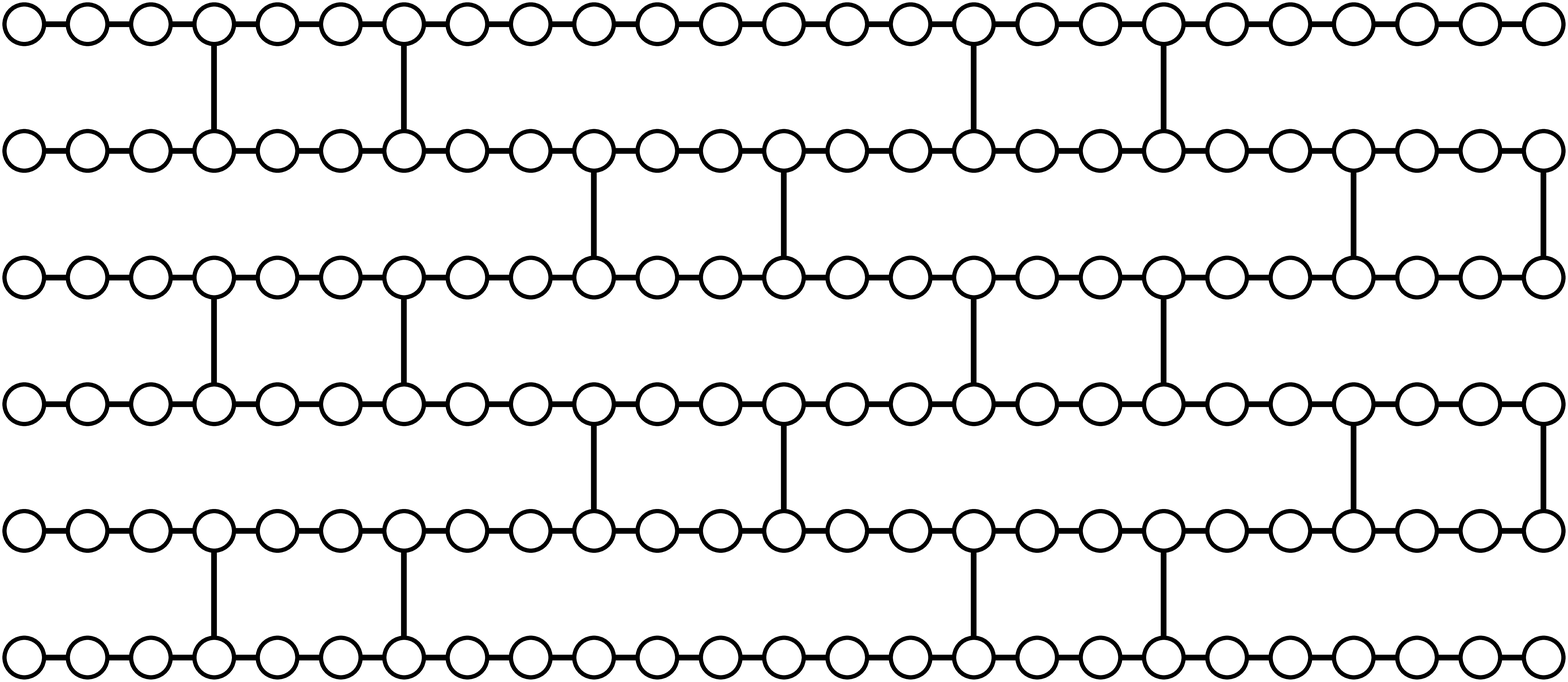}
\caption{
An example of the brickwork states.
}
\label{fig:bw}
\end{figure}
Below, we will show Problem~\ref{prob2} is BQP-hard.
We use the fact that approximation of
a matrix element $\bra{0}^{\otimes n}U\ket{0}^{\otimes n}$ 
of an $n$-qubit unitary circuit $U$ with an additive error $O(1/{\rm poly}(n))$ is BQP-hard~\cite{Bernstein1993,Neilsen2000,Arad2008}.
This is also the case when $U$ consists of 
a polynomial number of nearest-neighbor two-qubit gates 
acting on a one-dimensional array of qubits.
We have already established the relation between 
the partition function $Z_{G^{n \times m}}$ 
and the quantum circuit $\mathcal{C}$ as given in Eq.~\eqref{eq:hoge}.
Thus the goal here is to show that an arbitrary unitary circuit $U$
can be constructed by $\mathcal{C}$ with specific coupling strengths and magnetic fields.
This can be shown by using universality of MBQC on certain resource states
with a restricted type of projections,
which are available in Problem~\ref{prob2}.
A brickwork state~\cite{Raussendorf2005, Broadbent2009},
a type of graph states as shown in Fig.~\ref{fig:bw},
is useful for this purpose,
since we can show universality of MBQC on it
with a restricted type of single-qubit measurements.
Also in blind quantum computation,
the single-qubit measurements that Alice can command Bob to do in secrecy
is restricted.
Thus a brickwork state is utilized to show capability of universal blind quantum computation
using such a restricted type of measurements~\cite{Broadbent2009,Morimae2013}.

\begin{figure}
\centering
\includegraphics[width=85mm]{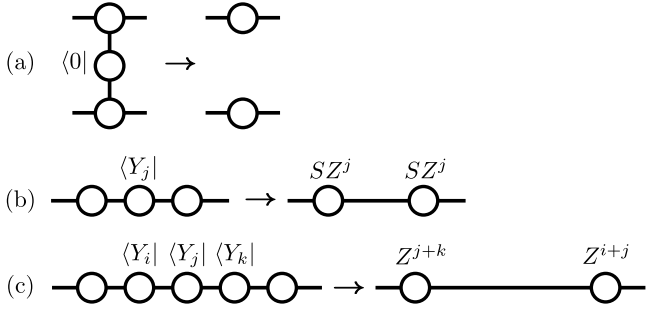}
\caption{
The rules for transformations of graph states 
by Pauli-basis projections.
(a) The projection $\bra{0}$ on a vertical edge qubit.
(b) The projection $\bra{Y_j}$ on an edge qubit.
The state $\bra{Y_j}$ ($j=0, 1$) is an eigenstate of the Pauli-$Y$ operator with an eigenvalue $(-1)^{j}$, i.e., 
$\bra{Y_j}\equiv(\bra{0}-(-1)^{j}i\bra{1})/\sqrt{2}$.
The local Clifford gate $SZ^j\otimes SZ^j$ is applied as a byproduct depending on the projected state $\bra{Y_j}$.
(c) A sequence of $Y$ projections connects 
the neighboring qubits directly up to the Pauli-$Z$ byproducts.
}
\label{fig:transformation rules}
\end{figure}

\begin{figure}
\centering
\includegraphics[width=85mm]{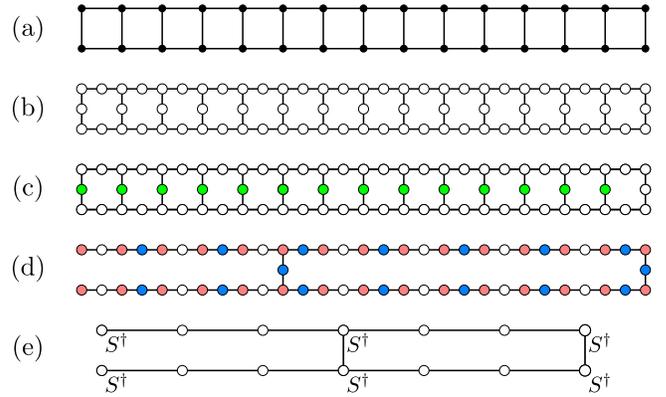}
\caption{(color online)
(a) The unit cell $G^{2\times 15}$.
(b) The corresponding decorated graph state $\ket{\tilde{G}^{2\times 15}}$.
(c) Those qubits colored by green are deleted by the projections $\langle 0|$.
(d)  Those qubits colored by red and blue,
corresponding to vertex and edge qubits, 
are projected by $\bra{Y_0}$ and $\bra{Y_1}$, respectively.
(e) After the projections, we obtain the 2D unit cell of 
the brickwork state up to local Clifford gates shown. 
}
\label{fig:brickwork unit}
\end{figure}

In order to obtain the brickwork state,
we transform the graph state by using the Pauli-basis projections
~\cite{Schlingemann2003, Hein2004}.
The transformation rules are summarized in Fig.~\ref{fig:transformation rules}.
For example, 
the $Z$-basis projection (with eigenvalue $+1$) removes the corresponding qubit 
the graph state
as shown in Fig.~\ref{fig:transformation rules} (a).
The $Y$-basis projection (with eigenvalue $+1$) connects the adjacent qubits directly
up to the local Clifford byproduct operator $S\otimes S$, as shown in Fig.~\ref{fig:transformation rules} (b), where $S=\mathrm{diag}(1,i)$.
Specifically, 
a sequence of the $Y$-basis projections on three neighboring qubits
connects the adjacent qubits directly 
up to the local $Z$ operator
as shown in Fig.~\ref{fig:transformation rules} (c).

We decompose $\bra{\alpha}$ as $\bra{\alpha}=\bra{\gamma}\otimes\bra{\delta}$,
where the projection
$\bra{\gamma}$ is used to transform the graph state $\ket{\tilde{G}^{n\times m}}$ into a brickwork state as follows.
Let us consider a unit cell $G^{2\times 15}$ of the square lattice $G^{n \times m}$ 
and the corresponding decorated graph state $|\tilde G\rangle$ as shown in Fig.~\ref{fig:brickwork unit} (a) and (b), respectively.
We perform the $\bra{0}$ projections on certain vertical edge qubits,
which are colored green in Fig.~\ref{fig:brickwork unit} (c), to cut the corresponding edges.
This is done by choosing the corresponding vertical coupling strengths to be $\beta J_{ab}^{\mathrm{v}} =0$. 
Next we perform the $Y$-basis projections on those qubits colored red
and blue in Fig.~\ref{fig:brickwork unit} (d).
Specifically, the red and blue colored qubits are
projected by $\langle Y_0| \equiv ( \langle 0 | -i \langle 1|)/\sqrt{2}$ and $\langle Y_1| \equiv ( \langle 0 | +i \langle 1|)/\sqrt{2}$,
respectively.
This is done by choosing the corresponding magnetic fields and coupling strengths to be $\beta h_a=i\pi/4$ and $\beta J^{\mathrm{h},\mathrm{v}}_{ab}=i\pi/4$ respectively.
By using the graph transformation rule shown in Fig.~\ref{fig:transformation rules},
we obtain a two-dimensional (2D) unit cell of the brickwork state as shown in Fig.~\ref{fig:brickwork unit} (e)
up to local Clifford gates.
If the above projections are made on the square lattice $G^{n\times m}$,
the brickwork state with local Clifford gates is prepared as shown in Fig.~\ref{fig:brickwork state} (c), 
which we define as $\ket{\mathrm{bw}}$.
At the top and bottom boundaries,
one-dimensional (1D) unit cells are also appeared.
(We can also consider a square lattice with a periodic boundary condition.)

These projections can be described as
\begin{equation}\label{eq:transformation to brickwork}
\braket{\alpha|{\tilde{G}^{n\times m}}}
=(\bra{\gamma}\otimes\bra{\delta})\ket{\tilde{G}^{n\times m}}
=\Delta_{\mathrm{t}}\braket{\delta|\mathrm{bw}}
\end{equation}
where $\Delta_{\mathrm{t}}=2^{-\#\gamma/2}$ with $\#\gamma$ being the number of the qubits in $\bra{\gamma}$, 
and $\bra{\delta}$ is a tensor product of the remaining horizontal edge qubits on the brickwork state.

The brickwork state $\ket{\mathrm{bw}}$ can be shown to be a universal resource for MBQC
using a restricted type of measurements available in Problem~\ref{prob2}:
\begin{lemm}[Universality of the brickwork state]
Let $U$ be an arbitrary quantum circuit 
consisting of a polynomial number of nearest-neighbor two-qubit gates in one dimension.
We can always find the horizontal coupling strengths $\beta J^{\mathrm{h}}_{ab} \in \{ i\pi/4 , \Omega\}$ for the unmeasured qubits, 
such that the projection $\bra{\delta}$ on them satisfies
\begin{align}
&\Abs{\bra{0}^{\otimes n}U\ket{0}^{\otimes n}-\Delta_{\mathrm{c}}^{-1}\braket{\delta|\mathrm{bw}}}\le \frac{1}{\mathrm{poly}(n)},
\end{align}
where $\Delta_{\mathrm{c}}=2^{-(\#\delta-n)/2}$ with $\#\delta$ being the number of the qubits in $\langle \delta |$.
\label{lem:universality}
\end{lemm}

\noindent{\it Proof of Lemma~\ref{lem:universality}:}
It is sufficient to show that a universal set of gates
can be implemented by choosing the remaining horizontal coupling strengths $\beta J^{\mathrm{h}}_{ab}$ from $\{
i \pi /4 , \Omega\}$.
The horizontal coupling strength
$\beta J^{\mathrm{h}}_{ab}=i \pi/4$ corresponds to 
the projections by $\bra{\alpha_{e_{ab}}}H= \langle +|S$, 
which results in $HS$ gate through gate teleportation.
Similarly,
the horizontal coupling strength
$\beta J^{\mathrm{h}}_{ab}=\ln(\sqrt{2}+1)/2+i \pi/4$ 
corresponds to 
the projections by $\bra{\alpha_{e_{ab}}}H= \langle + |T$ , 
which results in $HT$ gate through gate teleportation with $T=\mathrm{diag}(1,e^{i\pi/4})$.
[This can be confirmed by considering special instances
of the previous case as shown in Fig.~\ref{fig:algorithm for circuit} (d).]

\begin{figure*}
\centering
\includegraphics[width=170mm]{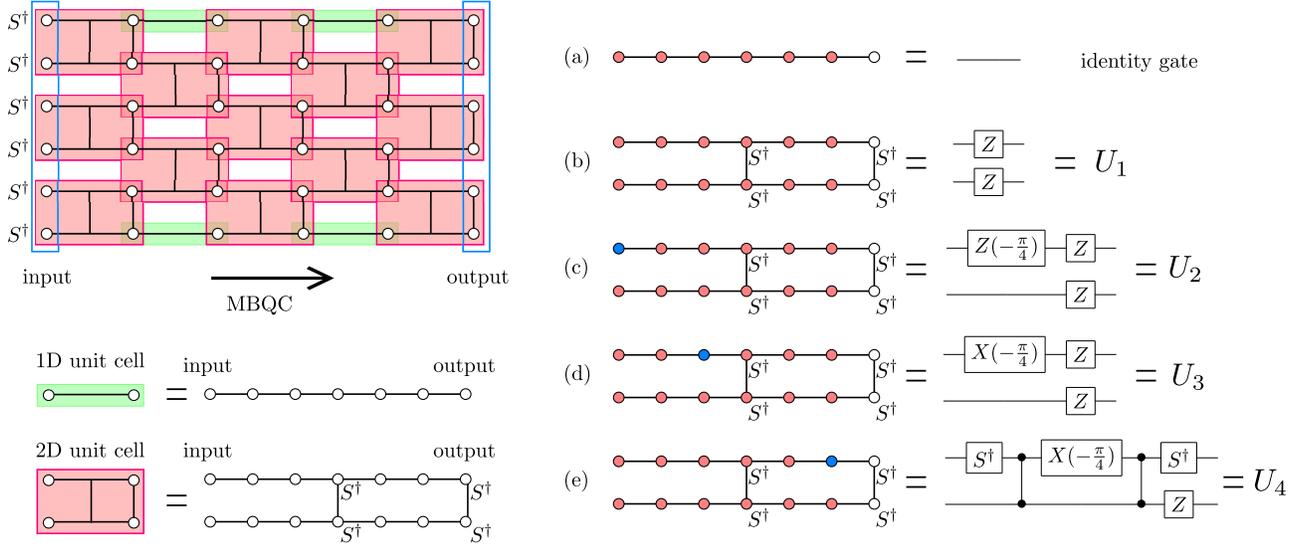}
\caption{
(color online)
(left) The MBQC interpretation 
of the projections on the brickwork state.
(right) The projections on the 1D and 2D unit cells
are translated into single- and two-qubit gates, respectively.
Those qubits colored by red 
are projected by
$\bra{Y_1}=\bra{+}S=(\bra{0}+i\bra{1})/\sqrt{2}$,
which corresponds to the coupling strength $\beta J_{ab}^{\mathrm{h}}=i\pi/4$. 
Those qubits colored by blue are projected by
$\bra{+}T=(\bra{0}+e^{i\pi/4}\bra{1})/\sqrt{2}$, 
which corresponds to the coupling strength $\beta J_{ab}^{\mathrm{h}}=\Omega$.
}
\label{fig:gate set}
\end{figure*}

As for the 1D unit cell of the brickwork state lying at the boundaries,
we can choose the coupling strengths such that the resulting single-qubit gate is an identity gate
as shown in Fig.~\ref{fig:gate set} (a).
Patterns of the horizontal coupling strengths 
for the 2D unit cell and 
the resulting two-qubit gates are shown in Fig.~\ref{fig:gate set} (b)-(e).
Specifically, the following two-qubit gates are realized:
\begin{align}
U_1&=Z\otimes Z,  \\
U_2&=(Z\otimes Z)\left(Z\left(-\frac{\pi}{4}\right)\otimes I\right),  \\
U_3&=(Z\otimes Z)\left(X\left(-\frac{\pi}{4}\right)\otimes I\right),  \\
U_4&=(S^\dag\otimes Z)\Lambda(Z)\left(X\left(-\frac{\pi}{4}\right)\otimes I\right)\Lambda(Z)(S^\dag\otimes I),
\end{align}
where $X(\theta) \equiv e^{-i\theta  X/2}$ and $Z(\theta)\equiv  e^{-i\theta  Z/2}$.
By using these two-qubit gates, the identity gate is constructed as $U_1{}^2$. The single-qubit $\pi/8$ gate is constructed as $T=(U_1U_2)^7$ up to a global phase.
$Z(k\pi/4)$ and $X(k\pi/4)$ gates for an integer $k=0,1,\dots,7$ are constructed as $Z(k\pi/4)=(U_1U_2)^{8-k}$ and $X(k\pi/4)=(U_1U_3)^{8-k}$, respectively. 
The Hadamard gate is constructed as $H=Z(\pi/2)X(\pi/2)Z(\pi/2)$.
The controlled-NOT gate is constructed as
\begin{widetext}
\begin{equation}
\Lambda_{2,1}(X)=
\left(X\left(\frac{\pi}{2}\right)\otimes I\right)
\left(Z\left(\frac{\pi}{2}\right)\otimes Z\left(-\frac{\pi}{2}\right)\right)
U_4U_1U_4
\left(Z\left(\frac{\pi}{2}\right)\otimes I\right).
\end{equation}
\end{widetext}
Since a universal gate set $\Set{T,H,\Lambda(X)}$~\cite{Neilsen2000}
is constructed,
the gate set $\Set{U_1,U_2,U_3,U_4}$ is also a universal set of gates.
This guarantees that the matrix element $\bra{0}^{\otimes n}U\ket{0}^{\otimes n}$
can be decomposed into a polynomial number of elementary gates $\Set{U_1,U_2,U_3,U_4}$
in the sense of an approximation by the Kitaev-Solovay algorithm~\cite{Neilsen2000}.
This yields
\begin{equation}
\left|\bra{0}^{\otimes n}U\ket{0}^{\otimes n} - \Delta_{\mathrm{c}}^{-1} \braket{\delta|\mathrm{bw}} \right| \leq \frac{1}{\mathrm{poly}(n)},
\end{equation}
where $\Delta_{\mathrm{c}}=2^{-(\# \delta -n)/2}$ meaning that
the probability amplitude is factored by $2^{-1/2}$
at each projections except for the projections for the final readouts of $n$ qubits.
\hfill $\Box$

By using Lemma~\ref{lem:universality}
and Eqs. \eqref{eq:partition function overlap} and \eqref{eq:transformation to brickwork},
we conclude that  
\begin{align}
\left|\Delta^{-1}Z_{G^{n\times m}} - \bra{0}^{\otimes n}U\ket{0}^{\otimes n} \right|
\leq \frac{1}{\mathrm{poly}(n)}
\end{align}
and $\Delta=\Delta_{\mathrm{c}}\Delta_{\mathrm{t}}\Delta_{\mathrm{o}}=2^{|V|/2+\#\Omega/4+n/2}$.
This indicates that
if we have an approximation $Z_{G^{n\times m}}^{\rm ap}$ of the partition function $Z_{G^{n\times m}}$ with an additive error 
$\Delta /{\rm poly}(n)$, then it satisfies
\begin{eqnarray}
\left|\Delta ^{-1} Z_{G^{n\times m}}^{\rm ap} - \bra{0}^{\otimes n}U\ket{0}^{\otimes n} \right| \leq \frac{1}{\mathrm{poly}(n)}.
\end{eqnarray}
This leads that Problem~\ref{prob2} is BQP-hard
and hence can simulate an arbitrary quantum computation.
This completes the proof of Theorem~\ref{the:hard}.

From Theorems \ref{the:algo} and \ref{the:hard},
we conclude that Problem~\ref{prob1} (and also Problem~\ref{prob2}) --
approximation of the partition functions of 
the Ising model on a square lattice 
with an additive error $\Delta /{\rm poly(n)}$--
is BQP-complete.

Problem~\ref{prob2} seems to be tight
in the sense that some coupling strengths are prohibited,
we could not show BQP-hardness.
If the horizontal coupling strength 
$\Omega$ is prohibited 
the corresponding quantum circuit $\mathcal{C}$ is a Clifford circuit,
and hence is classically simulatable~\cite{Neilsen2000}.
Similarly, if the vertical coupling strength $i\pi/4$ is prohibited,
the corresponding quantum circuit $\mathcal{C}$ is decomposed into
single qubit rotations without any interactions,
which apparently is classically simulatable.
If the vertical coupling strengths $\beta J_{ab}^{\mathrm{v}} =0$
is prohibited,
each qubit interact with nearest-neighbor qubits at every step.
Even in such a case,
there is a possibility to show universality
by using, for example, the scheme developed by Raussendorf~\cite{Raussendorf2005},
where spatially homogeneous operations with temporal modulations
are cleverly employed for universal quantum computation. 
BQP-hardness in such a case is an open problem for a future work.

Finally, we mention another subproblem of Problem \ref{prob1},
which can also be utilized to show BQP-hardness:
\begin{prob}[Another BQP-hard subproblem]
\label{prob3}
The horizontal coupling strengths 
are chosen to be $\beta J_{ab}^{\rm h}=i \pi/4$.
The vertical coupling strengths 
are chosen to be $\beta J_{ab}^{\rm h}= 0$ or  $=i \pi/4$.
The magnetic fields $\beta h_{a}$ are chosen from $\{ 0 , i\pi/4 ,i\pi/8\}$.
The problem is defined as approximation of 
the partition function $Z_{G^{n \times m}}$ of the given Ising Hamiltonian
$H_{G^{n\times m}}$ with an additive error 
$\Delta  /{\rm poly}(n)$,
where the approximation scale is defined to be $\Delta = 2^{n(m+1)/2}$.
\end{prob}
Problem \ref{prob3} is apparently a subproblem of Problem \ref{prob1}.
We can also show BQP-hardness of Problem \ref{prob3} straightforwardly
by following the strategy developed above.

\section{Extension to general coupling strengths and magnetic fields} \label{sec:IV}
In the previous section,
we have formulated the quantum algorithm to approximate 
the Ising partition functions
by using the overlap mapping and its MBQC interpretation 
reducing the approximation scale.
Unfortunately, the coupling strengths and magnetic fields in Problem~\ref{prob1}
take complex values.
In this section, we extend the domain of the proposed quantum algorithm
to general coupling strengths and magnetic fields,
including physical Ising models with real parameters,
which are of central interest in statistical physics
and computer science.
In such a case,
the projections are not always mapped into unitary quantum circuits,
but non-unitary operations appear.
Below we will first explain how to simulate
non-unitary operations originated from MBQC in the general parameter region
by introducing ancilla qubits. Based on this strategy,
the approximation scale $\Delta$ for the general domain is calculated.
We will confirm that the approximation scale
in the previous unitary case
can also be obtained as a special case.
If coupling strengths and magnetic fields are finite 
the approximation scale $\Delta$ for the extended version
is shown to be always smaller than 
that $\Delta _{\mathrm{o}}$ for the constant depth circuits
obtained solely from the overlap mapping.

Unfortunately,
we cannot show
BQP-hardness inside the physical region with
real coupling strengths and magnetic fields.
Thus it is still unknown 
the proposed algorithm does a nontrivial task inside this domain.
However, the extended quantum algorithm
also provides a partial evidence that
there is no efficient multiplicative approximation 
of the Ising partition functions with 
the real physical parameters.
This result strongly supports 
the proposed quantum algorithm does a nontrivial task
even in the physical parameter region.

\subsection{Simulation of linear operators}
We first explain how to simulate general liner operators
by using ancilla qubits and unitary gates,
following the scheme developed in Refs.~\cite{Aharonov2007,Arad2008}. 
Let $M$ be an arbitrary $d\times d$ matrix acting on a $d$-dimensional space.
The singular value decomposition yields
$M=WDP^{\dag}$, where $W$ and $P$ are unitary matrices.
$D = {\rm diag}(r_1, ..., r_d) $ is a diagonal matrix whose diagonal elements are real nonnegative values
being subject to $r_1\ge r_2\ge \dots\ge r_d\ge0$.
The eigenstate with the eigenvalue $r_i$ is denoted by $\ket{i}$ for all $i=1,2,\dots,d$.
If $d$ is finite, it is obvious that $W$ and $P^{\dag}$ can be implemented by a quantum computer.
Thus it is sufficient to consider implementation of the diagonal operator $D$ on a $d$-dimensional space.

In order to simulate $D$ we utilize an ancilla qubit $\ket{0}$ 
and a unitary operation $\tilde D$ on the composite system
\begin{align}
\tilde D &=\sum_{i=1}^d\ket{i}\bra{i}\otimes 
\Bigg[\frac{r_i}{r_1} \Big(\ket{0}\bra{0}+\ket{1}\bra{1}\Big)
\nonumber \\
&\qquad+\sqrt{1-\left(\frac{r_i}{r_1}\right)^2} \Big(-\ket{0}\bra{1}
 +\ket{1}\bra{0} \Big)\Bigg]
\nonumber \\
&\equiv
\sum_{i=1}^d\ket{i}\bra{i}\otimes Y(\theta_i),
\end{align}
where $Y(\theta _i)\equiv e^{-i\theta _i Y/2}$ is a $Y$-rotation gate on the ancilla qubit
with an angle $\theta_i\equiv 2\arccos\left(r_i/r_1\right)\in[0,\pi]$.
Since $Y(\theta_1)=I$, the unitary operator $\tilde{D}$ acts like a controlled-$Y$-rotation gate, which is controlled by the $d$-dimensional system.
Denoting the input state $\ket{\psi} \ket{0} = \left(\sum_{i=1}^{d}c_i\ket{i}\right) \ket{0}$,
this unitary operation yields
\begin{equation}
\tilde D \sum_{i=1}^{d}c_i\ket{i}\otimes\ket{0}=\sum_{i=1}^{d}c_i\ket{i}\otimes\left(\frac{r_i}{r_1}\ket{0}+\sqrt{1-\left(\frac{r_i}{r_1}\right)^2}\ket{1}\right).
\end{equation}
By projecting the ancilla qubit to $\ket{0}$,
we obtain
\begin{equation}
(I\otimes \bra{0}) \tilde D \sum_{i=1}^{d}c_i\ket{i}\otimes\ket{0}=\sum_{i=1}^{d}\frac{r_i}{r_1}c_i\ket{i}=\frac{1}{\Norm{M}} D \sum_{i=1}^{d}c_i\ket{i},
\end{equation}
where $\Norm{M} \equiv r_1$ is an operator one-norm.
Thus the linear operator $M$ is simulated as
\begin{equation}
(I\otimes \bra{0}) W \tilde D P ^{\dag} (\ket{\psi}\otimes\ket{0})=\frac{1}{\Norm{M}}M\ket{\psi},
\end{equation}
up to the factor $1/\Norm{M}$.
The simulation of linear operators
succeeds only when the ancilla qubit is projected 
by $\langle 0|$.
However, in the proposed quantum algorithm
no postselection is required,
since the matrix element 
of a unitary circuit including ancilla qubits is estimated
by using the Hadamard test as seen later.

\subsection{Extended quantum algorithm}
\begin{figure*}
\centering
\includegraphics[width=170mm]{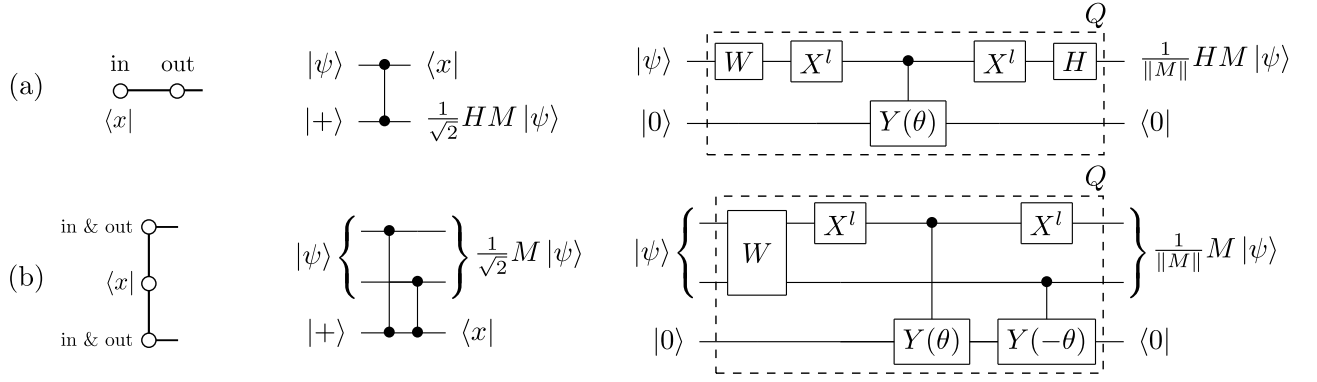}
\caption{
Projections on the graph states (left),
the resultant linear operations (middle), and 
the quantum circuits that simulate the corresponding linear operations (right).
The middle is a circuit representation of MBQC of the left
and the output is the r.h.s.\ of Eqs. \eqref{eq:general projection single} and \eqref{eq:general projection two}.
(a) The case of a projection on vertex and horizontal edge qubits, which we refer to as type I projection.
(b) The case of a projection on a horizontal edge qubit, which we refer to as type II projection.
}
\label{fig:general measurement}
\end{figure*}
Now we return to the quantum algorithm for the Ising models with the general parameter regions.
The projections in the MBQC interpretation of the overlap Eq.~\eqref{eq:partition function overlap}
can be classified into two, type I and II,
as depicted in Fig.~\ref{fig:general measurement} (a)
and (b), respectively.

Let us first consider a type I projection by $\bra{x}=x_0\bra{0}+x_1\bra{1}$,
where $x_0$ and $x_1$ are complex numbers with $|x_0|^2+|x_1|^2=1$.
For an arbitrary single-qubit input state $\ket{\psi}$,
this projection yields 
\begin{equation}\label{eq:general projection single}
(\bra{x}\otimes I)\Lambda(Z)(\ket{\psi}\otimes\ket{+})=\frac{1}{\sqrt{2}}HM\ket{\psi},
\end{equation}
where the resultant operator
\begin{equation}
M=\sqrt{2}\,\mathrm{diag}(x_0,x_1)
\end{equation}
is not a unitary gate in general [see Fig.~\ref{fig:general measurement} (a)]. 
The operator one-norm $\Norm{M}$ is given 
by $\Norm{M}=\sqrt{2}\max(|x_0|,|x_1|)$.
To simulate this operator $M$ on a quantum computer, 
we decompose it into $M=DW$, where $D$ is a positive diagonal operator and $W$ is a unitary operator,
\begin{equation}
D=\sqrt{2}\,\mathrm{diag}(|x_0|,|x_1|), \quad 
W=\mathrm{diag}(e^{i\phi_{x_0}},e^{i\phi_{x_1}}).
\end{equation}
Here $\phi_{x_{0,1}} = \arg x_{0,1}$ (i.e., $x_0=e^{i\phi_{x_0}}|x_0|$ and $x_1=e^{i\phi_{x_1}}|x_1|$).
We introduce an indicator function
\begin{equation}
l =
\begin{cases}
0&\text{if }\Abs{\frac{x_0}{x_1}}\ge 1, \\[6pt]
1&\text{if }\Abs{\frac{x_0}{x_1}}<1.
\end{cases}
\end{equation}
A $Y$-rotation gate is 
defined as 
\begin{align}
Y(\theta)&=
\begin{pmatrix}
\cos(\theta/2) & -\sin(\theta/2) \\
\sin(\theta/2) & \cos(\theta/2)
\end{pmatrix} \nonumber \\
&\equiv
\begin{pmatrix}
\Abs{\frac{x_1}{x_0}}^{(-1)^l} & -\sqrt{1-\Abs{\frac{x_1}{x_0}}^{2(-1)^l}} \\
\sqrt{1-\Abs{\frac{x_1}{x_0}}^{2(-1)^l}} & \Abs{\frac{x_1}{x_0}}^{(-1)^l}
\end{pmatrix},
\end{align}
where the angle $\theta\in[0,\pi]$ is given by
\begin{equation}
\theta = 2\arccos\left( \Abs{\frac{x_1}{x_0}}^{(-1)^l} \right).
\end{equation}
As previously mentioned,
we can simulate $M$ 
by using the controlled-$Y(\theta)$ gate $\Lambda(Y(\theta))$
as follows:
\begin{align}
\label{eq:general projection ancilla}
\frac{1}{\Norm{M}}HM\ket{\psi}
&=\bra{0}_2H_1X_1^l\Lambda_{1,2}(Y(\theta))X_1^lW_1\ket{\psi}_1\ket{0}_2
\nonumber \\
&\equiv\bra{0}_2 Q \ket{\psi}_1\ket{0}_2,
\end{align}
where the sequence of the unitary gates is denoted by
$Q$ [see the circuit diagram in Fig.~\ref{fig:general measurement} (a)].
From Eqs. \eqref{eq:general projection single} and \eqref{eq:general projection ancilla} we obtain
\begin{equation}
\bra{x}_1 \Lambda_{1,2}(Z)\ket{\psi}_1\ket{+}_2
=\frac{\|M\|}{\sqrt{2}} \bra{0}_2 Q\ket{\psi}_1\ket{0}_2.
\end{equation}

The type I projections are made 
on the vertex qubits and horizontal edge qubits.
Specifically, the projection $\bra{x}$ on the vertex qubit
is described as
\begin{equation}
\bra{x}=x_0\bra{0}+x_1\bra{1}=
\frac{e^{\beta h_a}\bra{0}+e^{-\beta h_a}\bra{1}}{\sqrt{\Abs{e^{\beta h_a}}^2+\Abs{e^{-\beta h_a}}^2}}.
\end{equation}
The indicator function, 
angle of the $Y$-rotation,
and norm are calculated for each vertex qubit $a\in V$
to be
\begin{align}
l_a &=\begin{cases}0&\text{if }\Abs{e^{2\beta h_a}}\ge1\\1&\text{if }\Abs{e^{2\beta h_a}}<1\end{cases},
\\
\theta_{a}&=
2\arccos\left( \left|e^{-(-1)^{l_a}2\beta h_a}\right| \right),
\\
\Norm{M_a} &=\frac{\sqrt{2}\Abs{e^{(-1)^{l_a}\beta h_a}}}{\sqrt{\Abs{e^{\beta h_a}}^2+\Abs{e^{-\beta h_a}}^2}}.
\end{align}
In the case of a horizontal edge qubit,
the projection is made with the Hadamard gate:
\begin{align}
\bra{x}&=x_0\bra{0}+x_1\bra{1}
=\frac{e^{\beta J_{ab}^{\mathrm{h}}}\bra{0}+e^{-\beta J_{ab}^{\mathrm{h}}}\bra{1}}{\sqrt{\Abs{e^{\beta J_{ab}^{\mathrm{h}}}}^2+\Abs{e^{-\beta J_{ab}^{\mathrm{h}}}}^2}}H .
\end{align}
Similarly to the previous case,
the indicator function, angle of the $Y$-rotation, and norm
are calculated for 
each horizontal edge qubit $\{a,b\}\in E^{\mathrm{h}}$
as follows:
\begin{align}
l_{{ab}} &=\begin{cases}0&\text{if }\left|\coth\left(\beta J_{ab}^{\mathrm{h}}\right)\right|\ge1\\1&\text{if }\left|\coth\left(\beta J_{ab}^{\mathrm{h}}\right)\right|<1\end{cases},
\\
\theta_{ab} &=
2\arccos\left( \left|\left\{\tanh\left(\beta J_{ab}^{\mathrm{h}}\right)\right\}^{(-1)^{l_{ab}}}\right| \right),
\\
\Norm{M_{{ab}}} &= \frac{\left|e^{\beta J_{ab}^{\mathrm{h}}}+(-1)^{l_{ab}}e^{-\beta J_{ab}^{\mathrm{h}}}\right|}{\sqrt{\left|e^{\beta J_{ab}^{\mathrm{h}}}\right|^2+\left|e^{-\beta J_{ab}^{\mathrm{h}}}\right|^2}}.
\end{align}

Next we consider the projection of type II
as shown in Fig.~\ref{fig:general measurement} (b).
A type II projection by $\bra{x}=x_0\bra{0}+x_1\bra{1}$ 
on an arbitrary two-qubit input state $\ket{\psi}_{1,2}$ yields
\begin{equation}\label{eq:general projection two}
(I_{1,2}\otimes\bra{x}_3)\Lambda_{2,3}(Z)\Lambda_{1,3}(Z)(\ket{\psi}_{1,2}\otimes\ket{+}_3)
=\frac{1}{\sqrt{2}}M\ket{\psi}_{1,2}
\end{equation}
where $M$ is given by
\begin{equation}
M=\mathrm{diag}(x_0+x_1,x_0-x_1,x_0-x_1,x_0+x_1),
\end{equation}
and $\Norm{M}=\max(\Abs{x_0+x_1},\Abs{x_0-x_1})$
[see Fig.~\ref{fig:general measurement} (b)].
The operator $M$ is decomposed into 
a positive diagonal operator $D$ and a unitary operator $W$:
\begin{align}
D&=\mathrm{diag}(\Abs{x_0+x_1},\Abs{x_0-x_1},\Abs{x_0-x_1},\Abs{x_0+x_1}), 
\\
W&=\mathrm{diag}(e^{i\phi_{x_0+x_1}},e^{i\phi_{x_0-x_1}},e^{i\phi_{x_0-x_1}},e^{i\phi_{x_0+x_1}}),
\end{align}
where $\phi_{x_0 \pm x_1} \equiv \arg (x_0 \pm x_1)$ (i.e., $x_0+x_1=e^{i\phi_{x_0+x_1}}\Abs{x_0+x_1}$ and $x_0-x_1=e^{i\phi_{x_0-x_1}}\Abs{x_0-x_1}$).
We introduce
an indicator function 
\begin{align}
l&=
\begin{cases}
0&\text{if }\Abs{\frac{x_0+x_1}{x_0-x_1}}\ge 1, \\[6pt]
1&\text{if }\Abs{\frac{x_0+x_1}{x_0-x_1}}<1.
\end{cases}
\end{align}
and a $Y$-rotation gate
\begin{align}
Y(\theta) &\equiv
\begin{pmatrix}
\Abs{\frac{x_0-x_1}{x_0+x_1}}^{(-1)^l} & -\sqrt{1-\Abs{\frac{x_0-x_1}{x_0+x_1}}^{2(-1)^l}} \\
\sqrt{1-\Abs{\frac{x_0-x_1}{x_0+x_1}}^{2(-1)^l}} & \Abs{\frac{x_0-x_1}{x_0+x_1}}^{(-1)^l}
\end{pmatrix}
\end{align}
where the angle $\theta\in[0,\pi]$
is given by 
\begin{equation}
\theta= 2\arccos\left(\Abs{\frac{x_0-x_1}{x_0+x_1}}^{(-1)^l}\right).
\end{equation}
Then, the linear operator $M$  
is simulated 
[see the circuit diagram in Fig.~\ref{fig:general measurement} (b)]
as follows:
\begin{align}
&\frac{1}{\Norm{M}}M\ket{\psi}_{1,2}
\nonumber \\
&=\bra{0}_3X^l_1\Lambda_{2,3}(Y(-\theta))\Lambda_{1,3}(Y(\theta))X^l_1W_{1,2}\ket{\psi}_{1,2}\ket{0}_3
\nonumber \\
&\equiv 
\bra{0}_3Q\ket{\psi}_{1,2}\ket{0}_3.
\label{eq:general projection ancilla 2}
\end{align}
From Eqs. \eqref{eq:general projection two} and \eqref{eq:general projection ancilla 2} we obtain
\begin{equation}
\bra{x}_3 \Lambda_{2,3}(Z)\Lambda_{1,3}(Z)\ket{\psi}_{1,2}\ket{+}_3
=\frac{\|M\|}{\sqrt{2}} \bra{0}_3 Q\ket{\psi}_{1,2}\ket{0}_3.
\end{equation}
[In a mild abuse of notation,
two types of circuits shown in 
Fig.~\ref{fig:general measurement} (a) and (b) are 
both denoted by $Q$. 
In the following,
the subscript $\eta$ of $Q_\eta$ identifies which of these two types of circuits is adopted.]

The type II projection corresponds to the projection on vertical edge qubits.
The projection for each vertical edge qubit is done by
\begin{align}
\bra{x}&=x_0\bra{0}+x_1\bra{1}=\frac{e^{\beta J_{ab}^{\mathrm{v}}}\bra{0}+e^{-\beta J_{ab}^{\mathrm{v}}}\bra{1}}{\sqrt{\Abs{e^{\beta J_{ab}^{\mathrm{v}}}}^2+\Abs{e^{-\beta J_{ab}^{\mathrm{v}}}}^2}}H  .
\end{align}
The indicator function, angle of the $Y$-rotation, and norm
are calculated for each vertical edge qubit 
$\{a,b\}\in E^{\mathrm{v}}$ as follows:
\begin{align}
l_{ab} &=  \begin{cases}0&\text{if }\left|e^{2\beta J_{ab}^{\mathrm{v}}}\right|\ge1\\1&\text{if }\left|e^{2\beta J_{ab}^{\mathrm{v}}}\right|<1\end{cases},
\\
\theta_{ab} &= 
2\arccos\left( \left|e^{-(-1)^{l_{{ab}}}2\beta J_{ab}^{\mathrm{v}}}\right| \right),
\\
\Norm{M_{{ab}}}&=\frac{\sqrt{2}\left|e^{(-1)^{l_{{ab}}}\beta J_{ab}^{\mathrm{v}}}\right|}{\sqrt{\left|e^{\beta J_{ab}^{\mathrm{v}}}\right|^2+\left|e^{-\beta J_{ab}^{\mathrm{v}}}\right|^2}}.
\end{align}

We have constructed 
quantum circuits $Q _{\eta}$ $(\eta \in V\cup E^{\rm h} \cup E^{\rm v})$ that simulate linear operators 
arising from the MBQC interpretation with the general parameters.
Each projection in 
the overlap mapping Eq.~\eqref{eq:partition function overlap}
is replaced with a unitary circuit $Q_\eta$,
as shown in Fig.~\ref{fig:general measurement} (right).
Including the initial state
and the final readout,
this yields
\begin{align}
Z_{G^{n\times m}}
&=\Delta_{\mathrm{o}}\braket{\alpha|\tilde{G}^{n\times m}}  \nonumber \\
&=\Delta \bra{0}^{\otimes |\tilde{V}|}\mathcal{C} \ket{0}^{\otimes |\tilde{V}|},
\end{align}
where the quantum circuit $\mathcal{C}$ 
is given by
\begin{eqnarray}
\mathcal{C} = \bigotimes _{a \in V_r} A_a  \left( \prod _{\eta \in \tilde V \backslash V_r}^{\to} Q_\eta  \right) \left( H^{\otimes n} \otimes I^{ \otimes |\tilde {V} - n|} \right).
\end{eqnarray}
The product
$\vec{\prod} _{\eta \in \tilde V\backslash V_r}$
is taken over all qubits on the decorated graph state $|\tilde G\rangle$
from the left to the right except for the vertex qubits at the right boundary.
The approximation scale is calculated to be
\begin{eqnarray}
\Delta = \Delta_{\mathrm{o}}
\prod_{v\in V \backslash V_r}\frac{\Norm{M_v}}{\sqrt{2}}
\prod_{e\in E^{\mathrm{h}}}\frac{\Norm{M_e}}{\sqrt{2}}
\prod_{e\in E^{\mathrm{v}}}\frac{\Norm{M_e}}{\sqrt{2}},
\end{eqnarray}
where the multiplication $\prod_{v\in V \backslash V_r}$ 
is taken except for the right boundary.

Similarly to the unitary case,
we can evaluate the matrix element $\bra{0}^{\otimes |\tilde{V}|}\mathcal{C} \ket{0}^{\otimes |\tilde{V}|}$
by using the Hadamard test.
Thus by using a quantum computer,
the partition function $Z_{G^{n\times m}}$ 
with the general coupling strengths and magnetic fields
can be approximated with an additive error $\Delta /\mathrm{poly}(n)$.
This concludes the extension of the algorithmic domain 
of the proposed quantum algorithm.

Let us discuss behavior of the approximation scale.
The norm $\Norm{M_\eta}$
is subject to $1 \leq \Norm{M_\eta} \leq \sqrt{2}$
for all $\eta \in V \cup E^{\rm h} \cup E^{\rm v}$.
If $\Norm{M_\eta} = 1 $ for all $\eta\in V \cup E^{\rm h} \cup E^{\rm v}$,
the multiplication of $\|M_\eta\|$ is the smallest.
In order to achieve this,
the coupling strengths and magnetic fields have to satisfy
\begin{align}
&\Re(\beta h_a) = 0,  \nonumber \\
&\Re(\beta J^{\mathrm{v}}_{ab}) = 0,  \\
&\Im(\beta J^{\mathrm{h}}_{ab})\in\Set{(2k+1)\pi/4|k\in\mathbb{Z}}, \nonumber 
\end{align}
where $\Re(\cdot)$ and $\Im(\cdot)$ indicate 
the real and imaginary parts respectively.
These conditions reproduce
the algorithmic domain and the approximation scale in the unitary case 
defined in Problem~\ref{prob1}.
This is because in the unitary parameter region in Problem~\ref{prob1},
the diagonal matrix $D$ becomes an identity, and hence
the angle $\theta$ of the $Y$-rotation is zero. 
This decouples the ancilla qubits
from the circuits. 
Then, the unitary gates $W$ and $H$, as shown in Fig.~\ref{fig:general measurement},
constitute the unitary circuit constructed in the previous section.
When the parameters are changed continuously, 
the approximation scale $\Delta$
is also changed continuously.
Thus we expect that an efficient 
approximation with this approximation scale 
is also hard for a classical computer
around the parameters in Problem~\ref{prob1}.

On the other hand, if $\Norm{M_\eta} = \sqrt{2}$ for all $\eta\in V \cup E^{\rm h} \cup E^{\rm v}$,
we obtain $\Delta_{\mathrm{o}} = \Delta$,
which means that the accuracy of approximation of the proposed quantum algorithm is 
equivalent to that of the constant depth algorithm mentioned in Sec.~\ref{sec:II}.
The conditions on the 
coupling strengths and magnetic fields read
\begin{align}
&\Re(\beta h_a)\to\pm\infty,  \nonumber \\
&\Re(\beta J^{\mathrm{v}}_{ab})\to\pm\infty,  \\
&\Re(\beta J^{\mathrm{h}}_{ab})=0\land \Im(\beta J^{\mathrm{h}}_{ab})\in\Set{k\pi/2|k\in\mathbb{Z}}. \nonumber 
\end{align}
If the parameters are chosen to be
finite, then 
the approximation scale $\Delta$ is
always smaller than that $\Delta _{\mathrm{o}}$ of the constant depth 
algorithm.
This indicates that the constructed quantum algorithm
does a better approximation than the constant depth algorithm
in almost all parameter region.

Let us examine
a representative example 
with $h_{a} =1$, $J^{\mathrm{v}}_{ab}=\pm 1$,
and $J^{\mathrm{h}}_{ab} =\pm 1$.
The partition function is given as a function of 
the inverse temperature, $Z(\beta)$.
The signs of the Ising interactions are chosen 
randomly with probability 1/2.
In this case, 
we can calculate the approximation scale explicitly
as follows:
\begin{equation}
\Delta =  
2^{nm} e^{(2nm-n-m)\beta} \left[\cosh(\beta)\right]^{nm-n}\left[\cosh(2\beta)\right]^{n/2}.
\end{equation}
Accordingly we can approximate the free energy per site
$F(\beta) = \ln Z(\beta) / (nm \beta )$
with an additive error 
\begin{eqnarray}
\epsilon (\beta) \equiv \ln( 1 + \Delta /[ {\mathrm{poly}(n)} Z(\beta)] )/(nm \beta).
\end{eqnarray}
Unfortunately, 
the approximation scale $\Delta$ still depends
on the size $n$ of the system. 
Thus an approximation of free energy per site 
with an additive error $1/{\rm poly}(n)$
cannot be achieved, although this is also the case for other 
quantum algorithms approximating the Ising partition functions~\cite{Master2003,Arad2008,Yung2010,Iblisdir2014}.

The accuracy 
of the proposed algorithm 
is comparable to that in Ref.~\cite{Iblisdir2014}
(at least in the size of the lattice mentioned),
which utilizes an analytical continuation 
in order to estimate the partition function with real parameters.
In the ferromagnetic case without magnetic fields,
the scheme in Ref.~\cite{Iblisdir2014} does
a better approximation at lower temperature.
This is because the scheme in Ref.~\cite{Iblisdir2014} 
intrinsically takes into account
the duality between low and high temperatures. 
On the other hand the proposed algorithm does not take into account it.
In general harder instances without any symmetry,
we expect that both schemes result in a comparable accuracy.

One advantage of the proposed algorithm is that
the approximation scale $\Delta$ 
can be calculated easily.
This property would be helpful to 
compare other approaches to approximate 
Ising partition functions.
Furthermore, the explicit construction of the unitary circuits
that approximate the Ising partition function with 
the physical parameter region also provides a 
clew to obtain a classical hardness result 
as discussed in the next subsection.

\subsection{A partial evidence of classical hardness of multiplicative approximation}
We have established 
a quantum algorithm that approximates the Ising partition functions
with the general coupling strengths and magnetic fields.
While Problem~\ref{prob1} has been shown to be BQP-complete,
it is still unknown whether or not the proposed quantum algorithm 
does a nontrivial task
in the physical parameter region with real coupling strengths and magnetic fields.
Thus there remains a possibility that a classical algorithm 
achieves a much better approximation in the physical parameter region. 
To reduce this possibility,
we show a partial evidence that 
an efficient multiplicative approximation cannot be attained 
by using a classical computer, unless the polynomial hierarchy 
collapses at the third level, which is highly implausible to occur.

Suppose we have a classical algorithm 
that approximates the Ising partition functions 
with an additive error:
\begin{eqnarray}
| Z_{G^{n \times m} } - Z_{G^{n \times m} }^{\rm ap} | \leq \frac{\epsilon \Delta }{\mathrm{poly}(n)}.
\end{eqnarray}
Here $\epsilon$ indicates the improvement 
made by the classical algorithm.
If $\epsilon \Delta /[Z_{G^{n \times m}}\mathrm{poly}(n)] \leq c$ with a constant $c$,
we can approximate the partition function with an multiplicative error
as follow:
\begin{eqnarray}
(1-c) Z_{G^{n \times m} }   \leq Z_{G^{n \times m} }^{\rm ap} \leq (1+c) Z_{G^{n \times m} }.
\end{eqnarray}
Below we will show a partial evidence that 
there is no classical algorithm that achieves an improvement $\epsilon$ such that $c \leq 1 - 2^{-1/4}$.
To this end, we show the following theorem
bridging the physical Ising partition functions and 
a class of quantum computation, so-called IQP~\cite{Bremner2010,Fujii2013}:

\begin{theo}[Ising partition functions and IQP]
The partition function $Z_{G^{n \times m} }$ of an Ising model on the square lattice $G^{n \times m}$
with real coupling strengths and magnetic fields is equivalent to 
a probability amplitude of an instance of IQP 
up to the scale factor 
\begin{align}
\Delta _{\mathrm{IQP}} &\equiv 
 \Delta_{\mathrm{o}}
2^{(|V|+|E^{\rm h}|)/2}
\prod_{v\in V }\frac{\Norm{M_v}}{\sqrt{2}}
\prod_{e\in E^{\mathrm{h}}}\frac{\Norm{M_e}}{\sqrt{2}}
\prod_{e\in E^{\mathrm{v}}}\frac{\Norm{M_e}}{\sqrt{2}}.
\end{align}
\label{IsingIQP}
\end{theo}

\noindent{\it Proof:}
Here we consider another quantum circuit 
\begin{eqnarray}
\mathcal{C}' =  \prod _{\eta \in \tilde V}^{\to} Q_\eta,
\end{eqnarray}
acting on $n+|\tilde V|$ qubits
where the initial and final state are 
$|+\rangle ^{\otimes n} |0\rangle ^{\otimes | \tilde V|}$ 
and $\langle +|^{\otimes n} \langle 0|^{ \otimes |\tilde V|}$,
respectively.
This quantum circuit also satisfies
\begin{eqnarray}
Z_{G^{ n \times m}} = \Delta' 
\langle +|^{\otimes n} \langle 0|^{ \otimes |\tilde V|} \mathcal{C}' 
|+\rangle ^{\otimes n} |0\rangle ^{\otimes | \tilde V|},
\end{eqnarray}
with the approximation scale
\begin{eqnarray}
\Delta' = \Delta_{\mathrm{o}}
2^{n/2}
\prod_{v\in V }\frac{\Norm{M_v}}{\sqrt{2}}
\prod_{e\in E^{\mathrm{h}}}\frac{\Norm{M_e}}{\sqrt{2}}
\prod_{e\in E^{\mathrm{v}}}\frac{\Norm{M_e}}{\sqrt{2}}.
\end{eqnarray}
(In contrast to the previous case,
the final projection is also simulated in $\mathcal{C}'$,
and hence the approximation scale 
$\Delta'$ is slightly different from $\Delta$.) 

\begin{figure*}
\centering
\includegraphics[width=170mm]{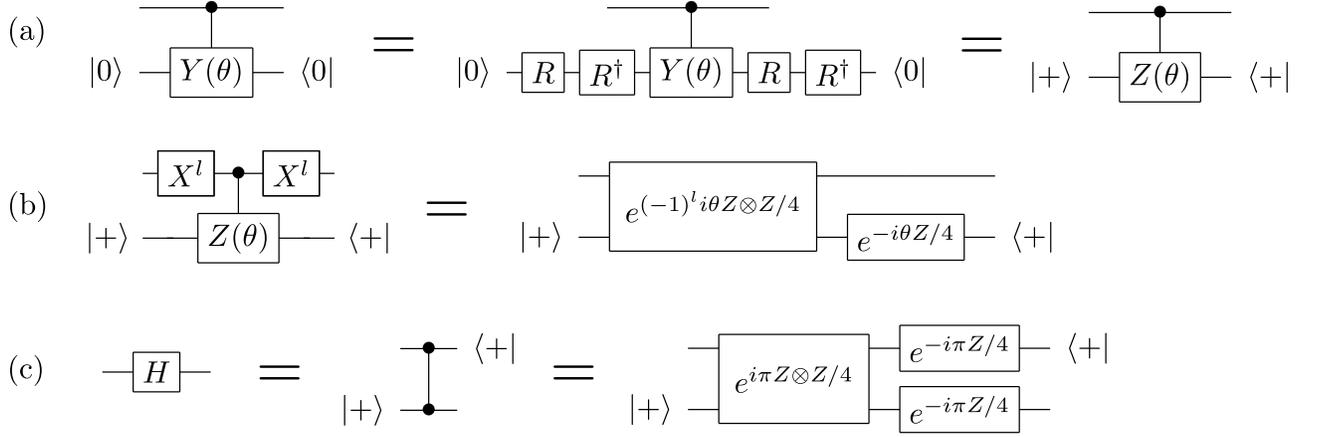}
\caption{
(a) A circuit equivalence between controlled-$Y$ and controlled-$Z$ rotation gates.
(b) Transforming a controlled-$Z$ gate to rotational gates with respect to the operators $Z$ and $Z\otimes Z$,
where $X^{l}$ is taken in the rotational angle.
(c) A measurement-based implementation of the Hadamard gate.
}
\label{fig:CY and H}
\end{figure*}

The quantum circuit $\mathcal{C}'$ consists of  
single-qubit gates $\{ X^{l}, H\}$, and two-qubit gates
$\Lambda (Y(\theta))$,
since $W$ becomes an identity gate in the physical parameter region.
By using a single-qubit Clifford gate $R=(X+Z+Y+iI)/2$,
the $Y$-rotation can be transformed into a $Z$-rotation $Z(\theta)=e^{ -i \theta Z/2}$ [see Fig.~\ref{fig:CY and H} (a)].
Then the initial and final states of the ancilla qubit
are transformed into $|+\rangle $ and $\langle +|$, respectively.
Thus we obtain
\begin{eqnarray}
Z_{G^{ n \times m}} = \Delta' 
\langle +|^{ \otimes n+|\tilde V|}  \mathcal{D}
 |+\rangle ^{\otimes n+| \tilde V|},
\end{eqnarray}
where $\mathcal{D}$ is obtained
from $\mathcal{C}'$ by replacing all controlled-$Y$-rotations $\Lambda (Y(\theta))$
to controlled-$Z$-rotations $\Lambda (Z(\theta))$.

The controlled-$Z$-rotation is decomposed into 
single- and two-qubit $Z$-rotations:
\begin{align}
\Lambda _{a,b} (Z(\theta))
&= e^{i\theta Z_a Z_b/4}e^{-i\theta Z_b/4}.
\end{align}
The two $X_a^l$ gates before and after $\Lambda_{a,b}(Z(\theta))$ 
are absorbed into the rotational angles [see Fig.~\ref{fig:CY and H} (b)]
\begin{align}
X_a^{l} \Lambda _{a,b} (Z(\theta))X_a^{l}
&= e^{(-1)^{l} i\theta Z_a Z_b/4} e^{-i\theta Z_b/4}.
\end{align}
The Hadamard gate can be implemented by using 
an ancilla qubit $|+\rangle$, two-qubit gate $\Lambda (Z)$,
and the projection $\langle +|$ in a teleportation-based way [see Fig.~\ref{fig:CY and H} (c)].
The $\Lambda (Z)$ gate can also be represented as single-
and two-qubit $Z$ rotations:
\begin{align}
\Lambda_{a,a'}(Z)
&= e^{-i\pi/4} e^{i\pi Z_a Z_{a'}/4} e^{-i\pi Z_a/4} e^{-i\pi Z_{a'}/4},
\end{align}
where the subscript $a$ and $a'$ denote the labels of the input and output qubits of the gate teleportation.

\begin{figure}
\centering
\includegraphics[width=85mm]{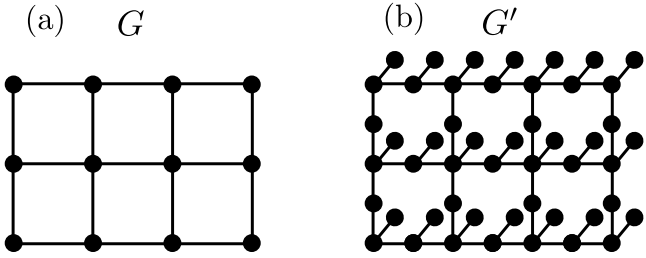}
\caption{
(a) The graph $G$, on which the Ising model
with real parameters is defined.
(b) The graph $G'$, on which the commuting circuits for IQP are defined.
The Ising model with imaginary parameters
is also defined on the graph $G'$.
}
\label{fig:G and G'}
\end{figure}

\begin{figure*}
\centering
\includegraphics[width=170mm]{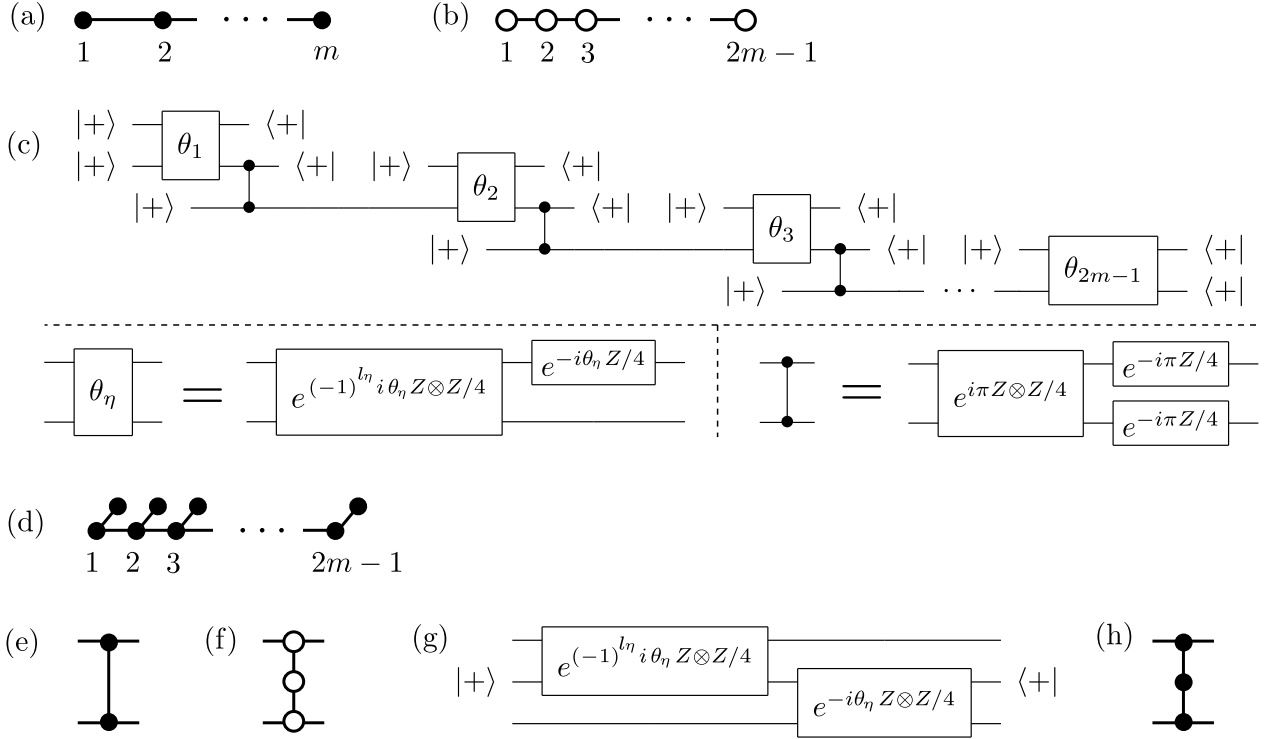}
\caption{
The correspondence between
the Ising models with real and imaginary parameters, 
on a square lattice $G$ and 
another lattice $G'$, respectively. 
(a)-(d) In the case of a 1D lattice with magnetic fields and horizontal couplings (i.e., $n=1$). 
(a) A 1D lattice $G$. (b) The corresponding decorated graph state $\ket{\tilde{G}}$.
(c) A quantum circuit $\mathcal{D}'$, which is decomposed into single- and two-qubit $Z$-rotation gates as shown in Fig.~\ref{fig:CY and H}.
(d) The graph $G'$, on which the Ising model with imaginary parameters is defined.
(e) A vertical coupling for the case of a square lattice $G$.
(f) The corresponding decorated graph state $|\tilde G\rangle$.
(g) The corresponding unitary gate in $\mathcal{D}'$. 
(h) The corresponding part of graph $G'$, on which
the Ising model with imaginary parameters is defined.
By combining (a)-(d) and (e)-(h)
we can obtain Theorem~\ref{IsingIQP}.
}
\label{fig:hoge}
\end{figure*}

Accordingly the circuit $\mathcal{D}$ can be reformulated as 
a commuting circuit $\mathcal{D}'$ acting on the qubits on a graph $G'$
as shown in Fig.~\ref{fig:G and G'}:
\begin{align}
& \langle +|^{ \otimes n+|\tilde V|}  \mathcal{D}
 |+\rangle ^{\otimes n+| \tilde V|}
\nonumber \\
&= 2^{(|V|+|E^{\mathrm{h}}|-n)/2}  \langle + |^{\otimes |\tilde V|+|V|+|E^{\mathrm{h}}|} \mathcal{D}' |+\rangle ^{\otimes |\tilde V|+|V|+|E^{\mathrm{h}}|},
\label{eq:IsingIQP}
\end{align}
where the final Hadamard gates are 
taken by the final state $\langle +|$
without teleportation,
and the number of qubits is equal to that of vertices $|V'|=|\tilde{V}|+|V|+|E^{\mathrm{h}}|$ of $G'$.
The commuting circuit $\mathcal{D}'$ consists only of single- and two-qubit $Z$-rotations
with appropriately chosen angles $\{ \tilde \theta _{ab}, \tilde \theta_{a} \}$ (see also Fig.~\ref{fig:hoge} and its caption):
\begin{align}
\mathcal{D}' &=
\prod_{\{ a,b\} \in E'} e^{i \tilde \theta_{ab}Z_a Z_b}
\prod_{a\in V'} e^{i\tilde \theta_a Z_a},
\end{align}
where the multiplication is taken over the set $E'$ of edges and $V'$ of vertices of the graph $G'$. 
The matrix element in the r.h.s.\ of Eq.~\eqref{eq:IsingIQP} is regarded as a probability amplitude of an instance of the IQP circuit.
Specifically the corresponding IQP circuit consists of
single-qubit and nearest-neighbor two-qubit commuting gates 
acting on a 2D graph $G'$.
Then we obtain the correspondence between
the Ising partition function and the probability amplitude of the IQP circuit,
\begin{align}
Z_{G^{n \times m} } = \Delta' 2^{(|V|+|E^{\mathrm{h}}|-n)/2} \langle + |^{ \otimes |V'|} \mathcal{D}' |+\rangle ^{ \otimes |V'|}.
\label{eq:IsingIQP2}
\end{align}
\hfill $\square$

Suppose the partition function $Z_{G^{n \times m} }$ can be approximated 
with a multiplicative error $2^{1/4}$, that is,
\begin{eqnarray}
2^{ -1/4} Z_{G^{n \times m} } \leq Z_{G^{n \times m} }^{\rm ap} \leq 2^{1/4} Z_{G^{n \times m} }.
\end{eqnarray}
Due to Theorem~\ref{IsingIQP}, this means that we can approximate the probability 
of the output of the corresponding IQP circuit
with a multiplicative error $\sqrt{2}$.
On the other hand,
as shown in Refs.~\cite{Bremner2010,Fujii2013},
even a weak simulation of a large class of IQP circuits with the multiplicative error $\sqrt{2}$ is hard for a classical computer unless the polynomial hierarchy 
collapses at the third level, which is highly implausible to occur.
A strong simulation, calculation of a probability distribution, is much harder than a weak simulation, 
which samples the outcomes according to the distribution.
Thus we reasonably conjecture that there is no efficient classical algorithm that approximates the Ising partition functions in the physical parameter region with a multiplicative error $2^{1/4}$.
(By considering a polynomial number of replicas of $Z_{G^{n \times m}}$,
the multiplicative error can be improved to be $2^{ 1/ {\mathrm{poly}}(n)}$,
although the following final result does not change.)
If this conjecture is true, 
the classical improvement of the approximation scale 
$\epsilon$ is limited to be
$\epsilon \geq (1- 2^{-1/4}){\rm poly}(n) Z_{G^{n \times m}}/\Delta$.
Since we are interested only in the exponential behavior,
a possible improvement of the approximation scale by a classical algorithm is 
$\epsilon \sim  Z_{G^{n \times m}}/\Delta$.

For the ferromagnetic Ising models with a constant magnetic field on arbitrary graphs, 
fully polynomial randomized approximation scheme (FPRAS) has been know to exist~\cite{Jerrum1993}.
However, under the random magnetic fields, approximation of ferromagnetic Ising 
partition functions belong, under an approximation-preserving reduction,
to a class \#BIS, which is defined as a counting problem of 
the number of independent sets of a bipartite graph~\cite{Goldberg2007}.
The class \#BIS is known to lie in-between FPRAS and \#SAT under an approximation-preserving reduction.
Here \#SAT indicates a counting problem of the number of satisfying configurations,
and does not 
have an efficient (polynomial) multiplicative approximation unless NP=PR~\cite{Zucherman}.
Moreover, it has been shown that a multiplicative approximation 
of antiferromagnetic Ising partition functions 
on $d$-regular graphs ($d \geq 3$) are NP-hard~\cite{SlySun}.
While an efficient approximation of Ising partition functions on the square lattices 
would still not be excluded,
these facts and the above partial evidence 
support a possibility that the proposed quantum algorithm of an additive approximation does a nontrivial task even in the physical parameter region.

There is also another interesting corollary of Theorem~\ref{IsingIQP}.
\begin{coro}[Real-imaginary correspondence]
\label{RealImagi}
An arbitrary Ising partition function $Z_{G^{n \times m}}$ on a square lattice $G^{n \times m}$
with real parameters can be mapped into 
an Ising partition function $Z_{G'}$ on a lattice $G'$ shown in Fig.~\ref{fig:G and G'} (b)
with imaginary parameters with a scale factor $\Delta' 2^{-5nm+2n+m}$:
\begin{eqnarray}
Z_{G^{n \times m}} = 
\Delta' 2^{-5nm+2n+m} Z_{G'}.
\end{eqnarray}
\end{coro}

\noindent{\it Proof:}
In Ref.~\cite{Fujii2013}, a correspondence between
IQP and Ising partition functions with imaginary parameters 
have been established.
It tells that the matrix element in the r.h.s.\ of Eq.~\eqref{eq:IsingIQP2}
is equivalent to an Ising partition function 
$Z_{G'}$ on a lattice $G'$ with imaginary parameters
with a scale factor $2^{|V'|}$:
\begin{equation}
\label{eq:pf on G'}
Z_{G'} =2^{|V'|} \langle + |^{\otimes |V'|} \mathcal{D}' |+\rangle ^{\otimes |V'|}.
\end{equation}
Combining Eq.~\eqref{eq:pf on G'} with Theorem~\ref{IsingIQP},
we obtain 
\begin{align}
Z_{G^{n \times m}} &= \Delta' 2^{(|V|+|E^{\mathrm{h}}|-n)/2}2^{-|V'|} Z_{G'}
\nonumber \\
&=\Delta' 2^{-5nm+2n+m} Z_{G'}.
\end{align}
\hfill $\square$

There has been a transformation, such as a duality transformation~\cite{Kramers}, that maps Ising partition function with a real coupling strength 
into an imaginary one
for a restricted case.
However, Corollary~\ref{RealImagi} can be applied for Ising models
with arbitrary real coupling strengths and magnetic fields.
Since imaginary and real Ising partition functions 
are well studied in quantum and classical information, respectively, 
the real-imaginary correspondence would be useful to 
bridge these two fields.

\section{Conclusions and discussions} \label{sec:V}
We have constructed a quantum algorithm for 
an additive approximation of the partition functions of 
Ising models on square lattices.
Specifically, we have argued both BQP-completeness~\cite{Nest2009,Cuevas2011}
and the extension toward the physical parameter region~\cite{Arad2008}
within the same model fixing the lattice geometry.
This allows us to calculate the approximation scale 
explicitly and to investigate the behavior of the approximation scale 
penetrating from the unitary case (Problem~\ref{prob1}),
which includes BQP-complete problem, to the physical parameter region,
which is of central interest in statistical physics and computer science.
We have shown that the MBQC interpretation
always provides a better approximation than 
the constant depth straightforward quantum algorithm
as long as the coupling strengths and magnetic fields are finite.

The overlap mapping and the MBQC interpretation 
are quite useful to translate the partition functions 
into quantum circuits and to calculate the resultant approximation scale.
While we have only considered square lattices,
this method could also be generalized to the Ising models on general lattice structures.
In such a case, the MBQC interpretation is made on general graph states.
In this context, Flow and its generalization, gFlow, theories~\cite{flow,gflow}
would provide an efficient scheme
to construct the corresponding quantum circuits.

Compared to the recent related work~\cite{Iblisdir2014}
based on an analytical continuation, 
the proposed construction with linear operator simulations 
provides a comparable approximation error
for the random-bond Ising models with magnetic fields
(at least with the size mentioned in Ref.~\cite{Iblisdir2014}).
One advantage of the proposed algorithm in the physical parameter region
is that the approximation scale can be easily obtained,
which allows us to compare the performance with other approaches.

We have also provided a partial evidence that 
there is no efficient classical algorithm 
for a multiplicative approximation of the Ising partition functions 
in the physical parameter region.
This has been shown by relating the quantum circuit 
that corresponds to the Ising partition functions to
an IQP circuit.
On the other hand, in the unitary case,
the problem (Problem~\ref{prob1}) that can be 
solvable by the proposed quantum algorithm 
is BQP-complete.
These facts strongly support that
the proposed quantum algorithm does a nontrivial task
even in the physical parameter regime with real coupling strengths and magnetic fields.

Unfortunately it is still unknown whether 
the proposed quantum algorithm does a nontrivial task
inside the physical parameter region.
However, the problems that we have to tackle are made clear now.
Firstly we have to rigorously proof that
classical simulation (weak simulation with a multiplicative error) of the related IQP circuits 
is hard.
This could be solved by clarifying whether or not
the IQP circuits become universal for quantum computation with the help of postselection~\cite{Bremner2010,Fujii2013}.
Secondary, we have to find a quantum algorithm or instances of the parameters that 
attain a multiplicative approximation.
Otherwise, we have to show that an additive approximation with the approximation scale $\Delta$
is still hard for a classical computer.
In doing so, the quantum circuits we have constructed to approximate the 
physical Ising model might provide us a clew.
If these would be accomplished,
we could have another nontrivial quantum algorithm
that solves quite important problems in statistical mechanics and computer science.

\begin{acknowledgements}
KF is supported by JSPS Grant-in-Aid for Research Activity Start-up 25887034.
This work was supported by JSPS Grant-in-Aid for Scientific Research(A) 25247068.

\end{acknowledgements}


\begin{thebibliography}{99}
\bibitem{RP=NP}
PR (randomized polynomial time computation) 
is a class of problems solvable by a probabilistic Turing machine
in the following sense: 
if the correct answer is NO, it always returns NO.
If the correct answer is YES, it returns YES with probability at least 1/2.
RP trivially includes P (polynomial time computation) and hence
RP=NP also implies P=NP, and hence is believed to be highly implausible.
\end{thebibliography}

\end{document}